\documentclass[usenatbib]{mn2e}
\usepackage[T1]{fontenc}
\usepackage{ragged2e}
\usepackage{graphicx}	
\usepackage{amsmath}	
\usepackage{amssymb}	
\usepackage{ae,aecompl}
\usepackage{xspace}
\usepackage{xcolor}
\usepackage{natbib}
\usepackage{epsfig}
\usepackage{txfonts}
\usepackage{pict2e}
\usepackage[colorlinks=true,allcolors=blue]{hyperref}
\usepackage{cleveref}

\usepackage{url}

\newcommand{\lsim}{\lesssim}

\newcommand{\expf}[1]{{{\rm e}^{#1}}}

\newcommand{\ion}[2]{{\text{{\sc #1}\,{\sc #2}}}}

\newcommand{\HeIlevel}[4]{{#1^{#2} {\rm #3}_{#4}}}   
\newcommand{\mHe}{{m_{\rm He}}}

\newcommand{\TCMB}{T_{\rm CMB}}

\newcommand{\xD}{{{x_{\rm D}}}}

\newcommand{\tauS}{{\tau_{\rm  S}}}

\newcommand{\tauct}{{\tilde{\tau}_{\rm  c}}}
\newcommand{\etac}{\eta_{\rm  c}}
\newcommand{\etact}{{\tilde{\eta}_{\rm  c}}}

\newcommand{\Planck}{{\sc Planck}}

\newcommand{\xe}{x_{\rm e}}

\newcommand{\id}{{\,\rm d}}

\newcommand{\beq}{\begin{equation}}   %

\newcommand{\eeq}{\end{equation}}   %

\newcommand{\beqa}{\begin{eqnarray}}   %

\newcommand{\eeqa}{\end{eqnarray}}   %

\newcommand{\beal}{\begin{align}}
\newcommand{\enal}{\end{align}}

\newcommand{\bspl}{\begin{split}}

\newcommand{\espl}{\end{split}}

\newcommand{\bsub}{\begin{subequations}}

\newcommand{\esub}{\end{subequations}}

\newcommand{\bmulti}{\begin{multline}}   %

\newcommand{\beqm}{\begin{mathletters}}   %

\newcommand{\eeqm}{\end{mathletters}}   %

\newcommand{\me}{m_{\rm e}}

\newcommand{\Te}{T_{\rm e}}

\newcommand{\sigT}{\sigma_{\rm T}}

\newcommand{\pot}[2]{#1 \times 10^{#2}}

\newcommand{\Yp}{Y_{\rm p}}

\newcommand{\omb}{\omega_{\rm b}}
\newcommand{\omc}{\omega_{\rm c}}

\newcommand{\thetaMC}{100\,\theta_{\rm MC}}
\newcommand{\ns}{n_{\rm s}}
\newcommand{\As}{A_{\rm s}}
\newcommand{\logA}{\ln\left(10^{10} A_{\rm s}\right)}
\newcommand{\aEMs}{\alpha_{\rm EM, 0}}
\newcommand{\mes}{m_{\rm e, 0}}
\newcommand{\aEM}{\alpha_{\rm EM}}

\newcommand{\LCDM}{$\Lambda$CDM\xspace}

\newcommand{\ho}{H_0}

\newcommand{\planck}{\emph{Planck}\xspace}
\newcommand{\Neff}{N_{\rm eff}}
\newcommand{\fede}{f_{\rm\, EDE}}

\newcommand{\zc}{z_{\rm\,c}}

\renewcommand{\xe}{X_{\rm e}}
\newcommand{\dxe}{\Delta \xe /\xe }

\newcommand{\vtwenty}{\emph{Voyage 2050}\xspace}
\newcommand{\vtwentyp}{\emph{Voyage 2050+}\xspace}
\newcommand{\vtwentypp}{\emph{Voyage 2050++}\xspace}
\newcommand{\spixie}{\emph{SuperPIXIE}\xspace}
\usepackage{newtxtext,newtxmath}

\newcommand{\changeJ}[1]{{#1}}
\newcommand{\changeL}[1]{{#1}}

\title[Probing fundamental physics with the CRR]{Using the cosmological recombination radiation to probe early dark energy and fundamental constant variations}

\author[L.~Hart and J.~Chluba]{
Luke Hart$^{1}$ and Jens Chluba$^{1}$
\\
$^{1}$Jodrell Bank Centre for Astrophysics, Alan Turing Building, University of Manchester, Manchester M13 9PL}

\date{\vspace{-5mm}Accepted XXX. Received YYY; in original form ZZZ}

\pubyear{2022}
\begin{document}
\label{firstpage}
\pagerange{\pageref{firstpage}--\pageref{lastpage}}

\maketitle

\begin{abstract}
The cosmological recombination radiation (CRR) is one of the guaranteed spectral distortion signals from the early Universe. The CRR photons from hydrogen and helium pre-date the last scattering process and as such allow probing physical phenomena in the pre-recombination era.
Here we compute the modifications to the CRR caused by early dark energy models and varying fundamental constants. These new physics examples have seen increased recent activity in connection with the Hubble tension, motivating the exploratory study presented here. The associated CRR responses are spectrally-rich but the level of the signals is small. %
We forecast the possible sensitivity of future spectrometers to these effects. 
Our estimates demonstrate that the CRR directly depends to changes in the expansion history and recombination physics during the pre-recombination era. However, futuristic sensitivities are required for spectrometer-only constraints that are competitive with other cosmological probes.
Nevertheless, measurements of the CRR can directly reach into phases that otherwise remain inaccessible, highlighting the potential these types of observations could have as a probe of the early Universe. 
A combination with \planck data further shows that a synergistic approach is very promising.
\end{abstract}


\section{Introduction} 
\label{sec:intro} 
In modern cosmology, our detailed understanding of the cosmic microwave background (CMB) has opened the floodgates to precision tests of the \LCDM model and different flavours of new physics beyond. This has been made possible with precise measurements of the CMB anisotropies using space- and ground-based experiments \citep{wmap9results, ACTPol, Planck2018over}. Next generation CMB experiments furthermore promise to uncover unparalleled details of the background radiation, allowing us to probe even deeper into the underpinnings of cosmology \citep{Matsumura2014, CMBS42016, SOWP2018}. 

In spite of the great successes for the $\Lambda$CDM model, many extensions have been considered. 
Additions to the standard model of cosmology have included {\it modifications of the effective numbers of relativistic species} and {\it sterile neutrinos} \citep{PlanckNeutrino, Nollett2011, BattyeNeutrinos, Lesgourgues2014, Abazajian2015}, {\it dark matter annihilation} \citep{Padmanabhan2005, Galli2009, Huetsi2009, Chluba2010a}, {\it primordial magnetic fields} \citep{Sethi2005, Shaw2010PMF, Kunze2014, Chluba2015PMF, Planck2016PMF, Paoletti2019, Jedamzik2019, Jedamzik2020} and {\it variations of fundamental constants} \citep{Avelino2001,Battye2001, Galli2011,Menegoni2012,Planck2015var_alp, Hart2017, Hart2020a}. So far, no {\it significant deviation} from the \LCDM model has been identified.
However, several tensions have been discussed \citep[see][for review]{Perivolaropoulos2021, Abdalla2022tensions}. Of these, the \emph{Hubble tension}, a discrepancy of the average expansion rate between the early and late Universe, seems to persist and grow in significance, suggesting modifications to fundamental physics might be required \citep{Bernal2016, Verde2019, Divalentino2020Intertwined}.

Indeed, several of the aforementioned new physics examples have been applied in an attempt to alleviate the notorious Hubble tension \citep[see][for a comparative study]{Schoeneberg2021}. 
One viable solution is connected to the presence of an oscillating scalar field, such as those relating to \emph{ultra-light axions} (henceforth ULA), as has been considered with the consequences for the CMB anisotropies outlined in \citet{Poulin2018}. The dynamical dilution of this field in the pre-recombination era leads to an \emph{early dark energy} phenomenon, changing the expansion rate before last scattering \citep{Poulin2019}. This scalar field can be treated like an effective dark energy fluid, an approximation which has been rigorously tested against the direct field evolution \citep{Smith2020}. However, the jury is still out on whether or not this can be reconciled with large-scale structure and BAO data at lower redshifts \citep{Ivanov2020EDE,Smith2020LSS, Simon2022, Cruz2022}. Several similar dark energy theories have built on these ideas \citep[e.g.,][]{Lin2019, Alestas2020, Hill2020EDE,McDonough2021, Karwal2022, Wang2022,Kojima2022}, and most recently, even a link to possible detection of birefringence was drawn \citep{Murai2022}.

Another promising addition to the standard model of cosmology that could alleviate the Hubble tension involves the variations of fundamental constants \citep{Hart2020a}. In the interactions between matter and radiation, the main fundamental constants that garner interesting physical insights are the fine structure constant ($\aEM$) and the effective electron mass ($\me$). 
At low redshifts ($z\lesssim 2$), the fine-structure constant has been tested with many astrophysical probes such as quasar absorption spectra \citep{Bize2003, Murphy2017, Kotus2017, Levshakov2019, Wilczynska2020VFC}, white dwarves \citep{Hu2020VFC} and black holes \citep{Hees2020VFC}.
More recently, variations in the electron-proton mass ratio have also been studied using quasar spectra \citep{Levshakov2020_me}. These works all indicate consistency with the standard value known from local lab experiements.

At higher redshifts, the changes in the 21cm line radiation that arise due to variations in the fine structure constant have been forecasted for future experiments \citep{Lopez2020}. Similarly, the changes to the light element abundances arising from Big Bang Nucleosynthesis (BBN) have been tested for fundamental constant variations \citep{Avelino2001,Coc2013,Alvey2019}. 
However, the effects of varying fundamental constants (VFCs) on the CMB anisotropies indicate an interesting avenue in connection with the Hubble tension. With the most recent \Planck data, the specific dependencies of these constants during recombination have shown unique imprints during hydrogen and helium recombination \citep{Hart2017}. Specifically, the variations from the recombination epoch lead to a significant geometric degeneracy between $\me$ and $H_0$ which can alleviate the Hubble tension \citep{Hart2020a}.
Several reviews have been published on the motivation and various methods of detecting VFCs \citep{Uzan2003, Uzan2011, Martins2017review}, and it is important to ask if there are indeed new methods for shedding light on early VFCs. 

The study of models that alter the expansion history of the universe can be carried out with many of the aforementioned cosmological probes (e.g., CMB anisotropies, weak lensing, 21cm). As explored recently, primordial $\mu$-type spectral distortions of the CMB may also provide information on the expansion rate  \citep{Lucca2020b}. However, a particular distortion that directly probes different periods of cosmological time is the \emph{cosmological recombination radiation} (CRR) from $z\simeq 1000-8000$ \citep{Sunyaev2009}. Predicted by \LCDM, this distortion arises as the CMB photon field departs from thermal equilibrium due to the transitions within hydrogen and helium atoms as well as the continuum during recombination \citep{Zeldovich68,Peebles68}. This manifests as a unique spectral signal in the CMB spectrum \citep{Dubrovich1975, RybickiDell94}, which can now be accurately computed using {\tt CosmoSpec} \citep{Chluba2016CosmoSpec}. Given the superposition of hydrogen and helium lines, the spectral changes caused by variations in cosmological parameters can be constrained with futuristic spectrometers (e.g., \emph{PRISM}, \vtwenty) \citep{PRISM2013WP,PRISM2013WPII,Vince2015, Mayuri2015, Chluba2019Voyage,Hart2020c}. 
Similarly, new physics can affect the dynamics of the recombination process and thus leave unique imprints in the CRR \citep{Jose2008, Chluba2008T0, Chluba2008c}.

In this paper, we will outline the ways that the CRR can probe the effects from early dark energy theories and variations of the fundamental constants $\aEM$ and $\me$. In Section~\ref{sec:ede}, we introduce the approach used in the previous ULA constraints papers \citep{Poulin2018, Poulin2019} and revisit the main effects on the background expansion history. We then study how early dark energy can affect the ionisation history and consequently, impact the recombination lines in unique ways. 
We briefly discuss the detectability of these variations for different models using rudimentary signal-noise predictions and then investigate more complete parameter correlations with a Fisher matrix analysis. 
Our estimates are meant to give a first rough feeling about the observability of these effects; however, a rigorous analysis in combination with CMB anisotropy constraints is left for a future investigation.

In Section~\ref{sec:vfc}, we show the differences in the CRR caused by VFCs. We explain how these variations can be related to the features discussed in \citet{Hart2017} with a particular emphasis on the modifications to the recombination process caused by these changes. 
We provide a comparative study indicating the impact of including CMB anisotropy results with future spectrometers. This leads to discussion on the possible solutions to the Hubble tension involving $\me$ \citep{Hart2020a,Hart2021}.

\section{Early dark energy}\label{sec:ede}
The equations of motion for the ULA can be reconstructed using the effective fluid approximation as has been validated in comparison to the full scalar field evolution \citep{Smith2020}. The field dynamics lead to an evolving energy density, 
\begin{equation}
    \Omega_\phi\left(z\right) = \frac{2\Omega_\phi\left(\zc\right)}{1+\left[\left(1+\zc\right)/\left(1+z\right)\right]^{\,3\left(1+w_n\right)}}\;,
\end{equation}
with an equation of state,
\begin{equation}
    1+w_\phi(z) = \frac{1+w_n}{1+\left[\left(1+z\right)/\left(1+\zc\right)\right]^{\,3\left(1+w_n\right)}}\;.
\end{equation}
Here $\zc$ signifies the redshift when the field becomes dynamical and $n$, the order of the oscillating potential for a ULA field \cite[][for more details]{Poulin2018}, determines the dilution rate of the energy density according to $w_{n} = \left(n-1\right)/\left(n+1\right)$.

\begin{figure}
    \centering
    \includegraphics[width=.98\columnwidth]{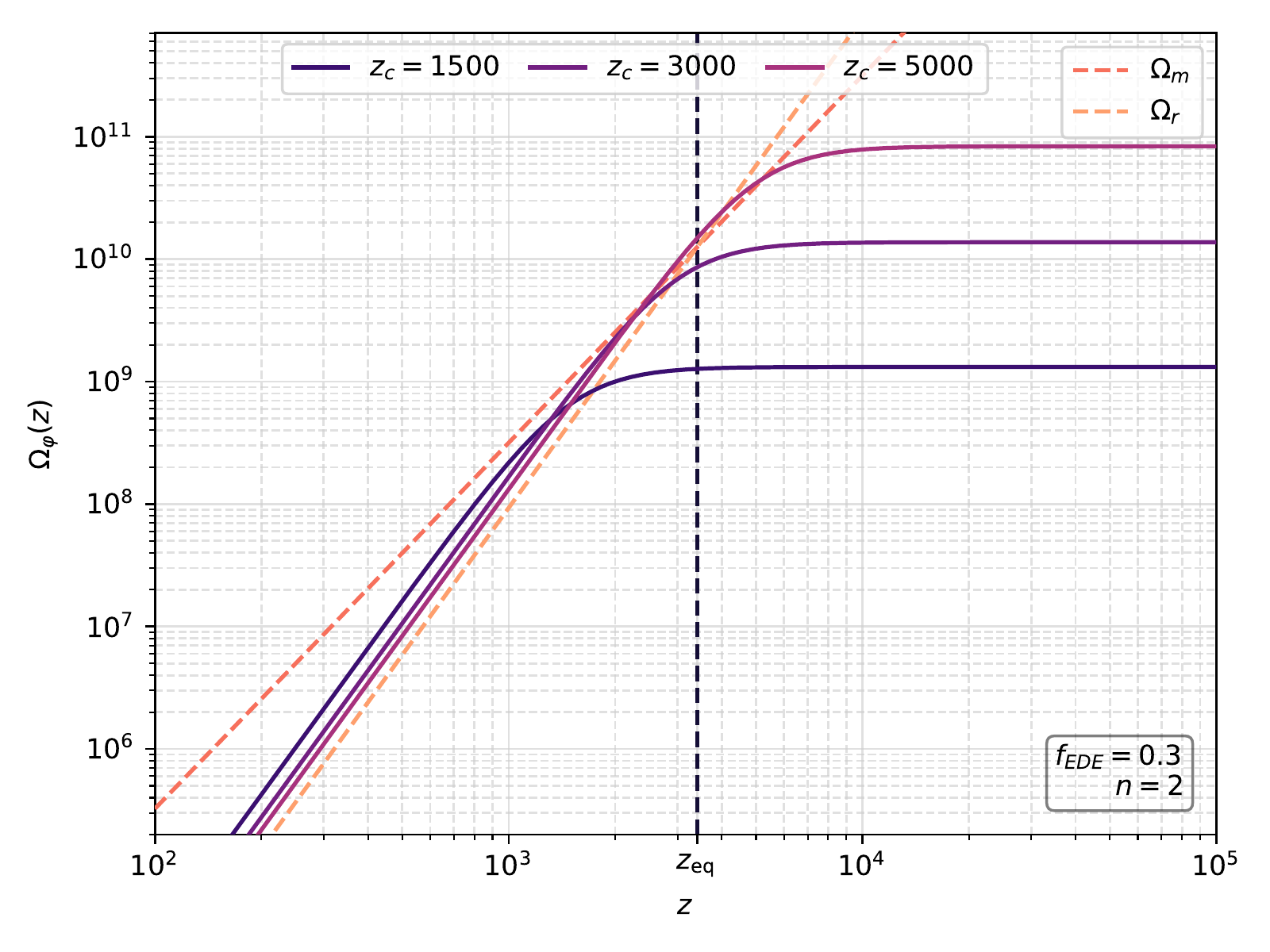}
    \\
    \includegraphics[width=.98\columnwidth]{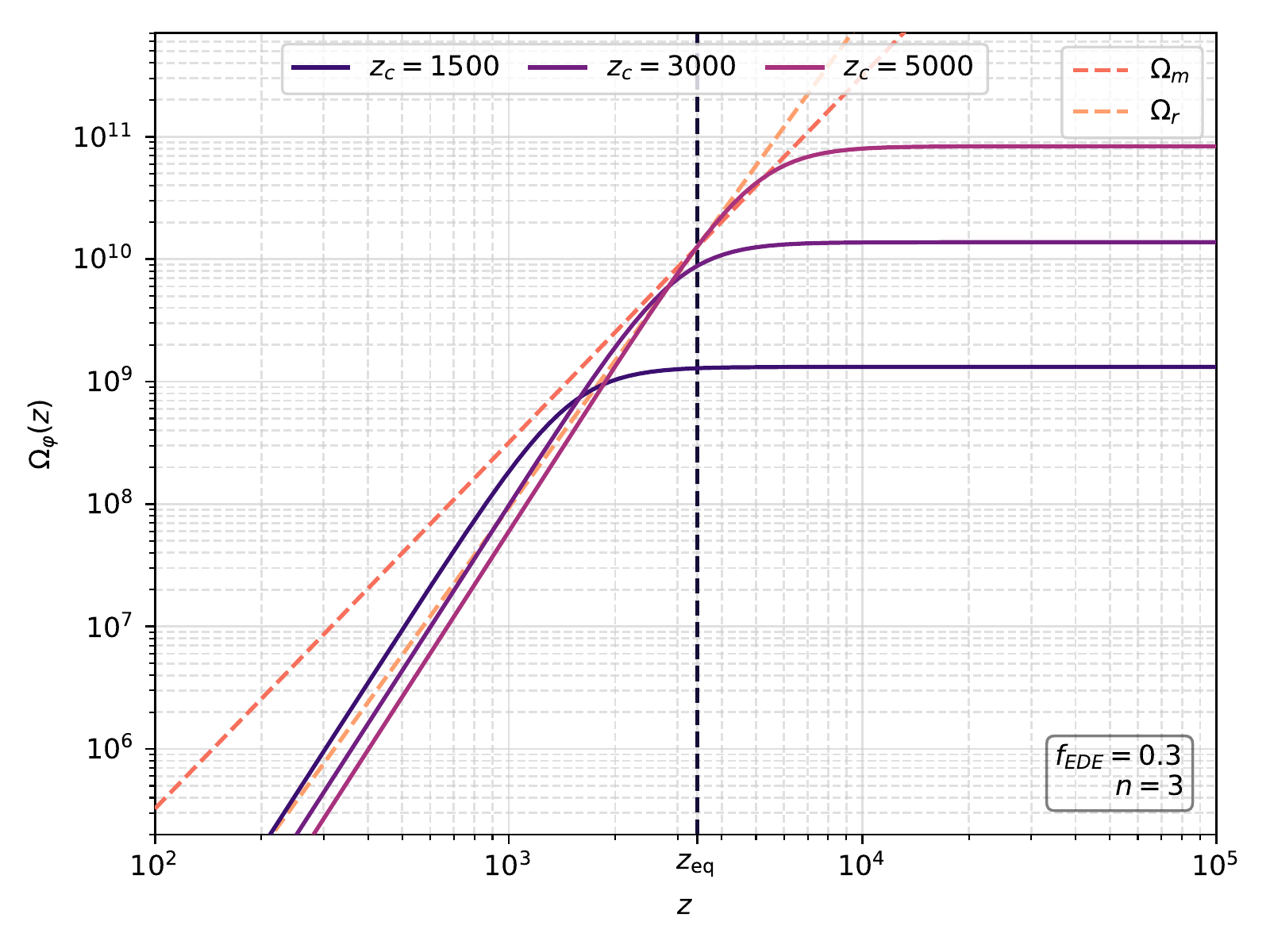}
    \\
    \includegraphics[width=.98\columnwidth]{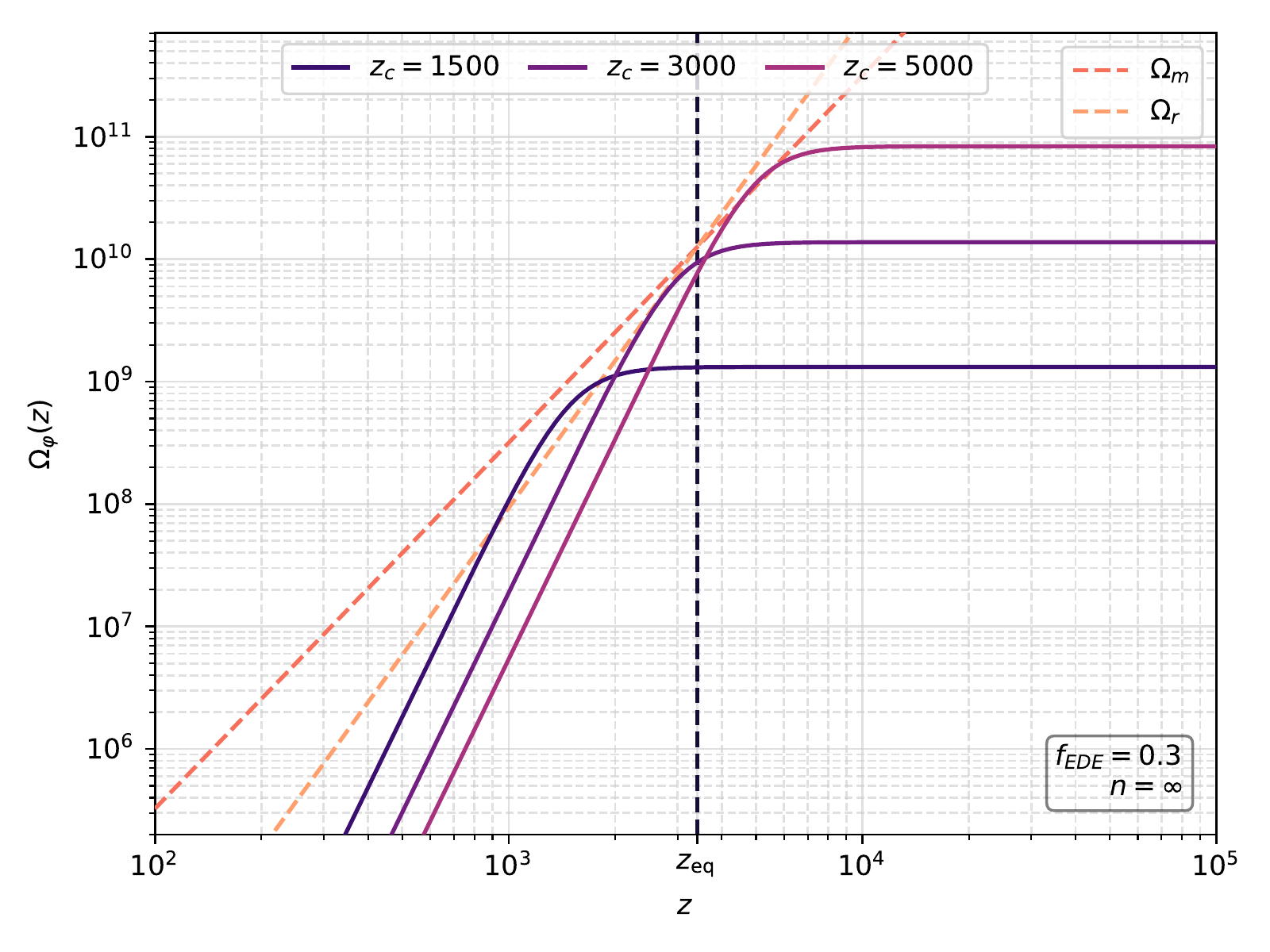}
    \\
    \caption{The energy density for EDE under the fluid-approximation for $\fede=0.3$. Curves for 3 different critical redshifts $\zc$ are shown for $n=2$ ({\it top}), $n=3$ ({\it middle}) and $n=\infty$ ({\it bottom}) panels, respectively. The matter and relativistic density evolution are shown for comparison as well as the matter-radiation equality epoch.}
    \label{fig:ede}
\end{figure}

The energy density for several potential orders, $n$, are illustrated in Fig.~\ref{fig:ede}. To quantify the amplitude of the early dark energy density, we will use the parametrisation $\fede = \Omega_\phi(\zc)/\Omega_{\rm tot}(\zc)$, following previous ULA papers. At early times ($z\gg \zc$), the early dark energy behaves like a cosmological constant due to the \emph{Hubble friction} term of the evolving field. As the dark energy fluid becomes dynamical, it decays according to $\Omega_{\phi}\propto\left(1+z\right)^{3\left(1+w_n\right)}$. For $n=2,3$ this means the early dark energy density drops off like radiation or an ultra-relativistic species respectively. In the extreme case that $n\rightarrow\infty$, the field energy density rapidly decays $\propto a^{-6}$. This corresponds to the field energy being totally dominated by kinetic term, similar to a scalar field term\footnote{This was explored with detailed changes to the background cosmology in \citet{Karwal2016}.}.

\begin{figure}
    \centering
    \includegraphics[width=\columnwidth]{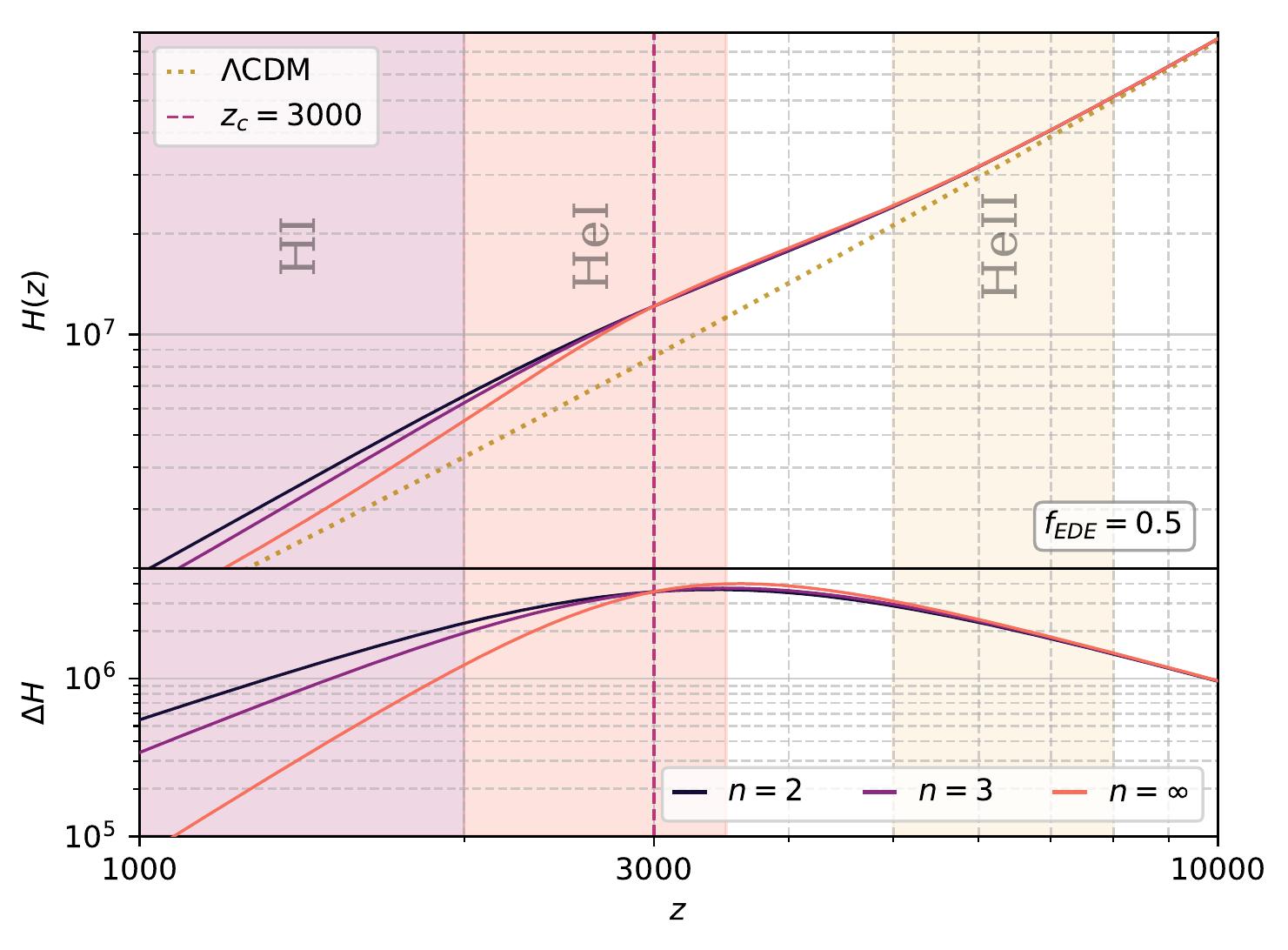}
    \\
    \caption{Examples of the evolution of the Hubble parameter $H(z)$ as a function of the different indices $n=\{2,3,\infty\}$ at $\zc=3000$. The parameter growth with comparison against \LCDM (gold, dashed) is shown in the top panel, while the difference of the models, $\Delta H$, are shown in the bottom panel. Overlaid are the rough neutral hydrogen, neutral and ionised helium recombination eras \citep[see][for more details]{Sunyaev2009}. The Hubble parameter here is measured in ${\rm km \,s^{-1} Mpc}$, as usual.}
    \label{fig:hz}
\end{figure}

\begin{figure}
    \centering
    \includegraphics[width=\columnwidth]{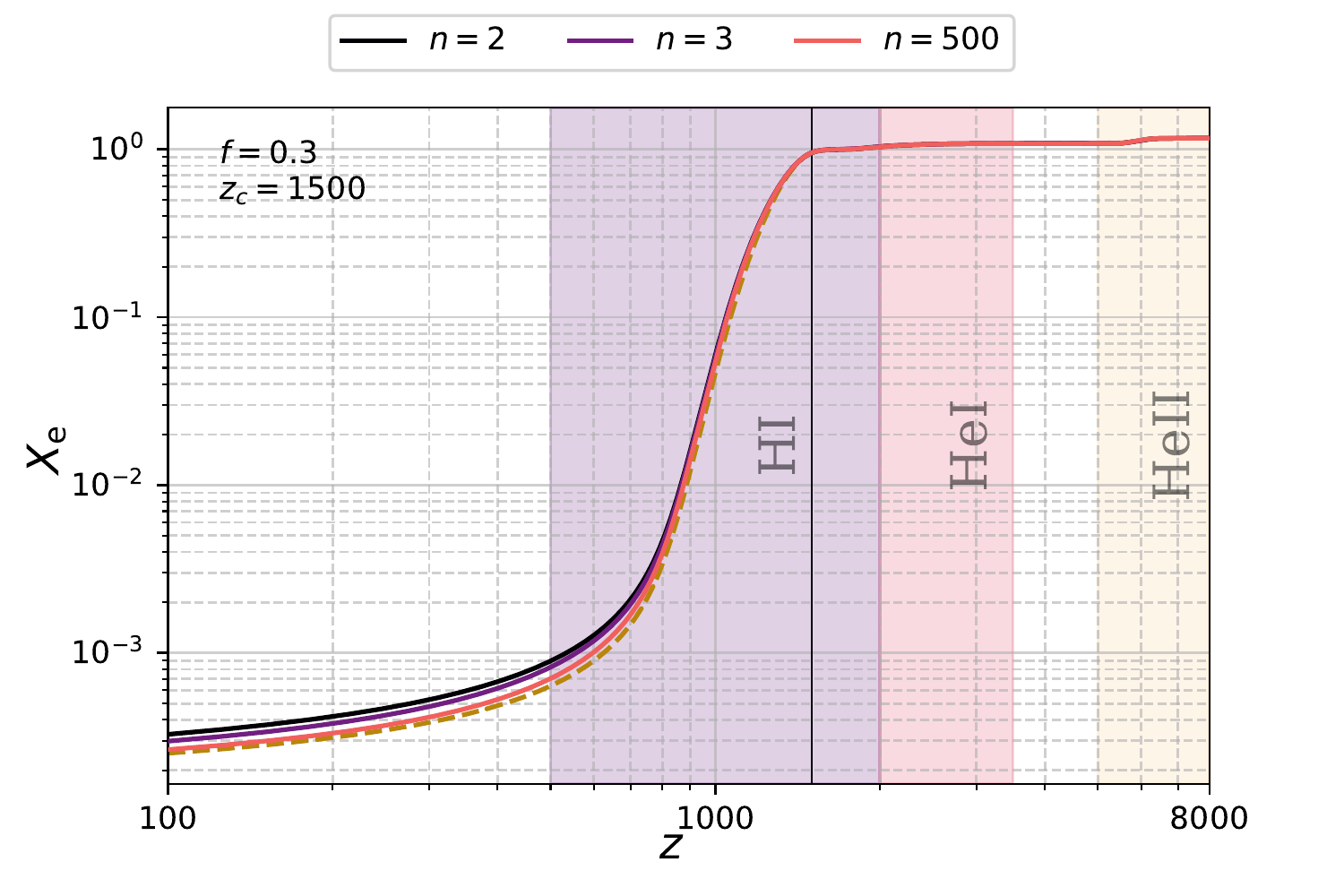}
    \includegraphics[width=\columnwidth]{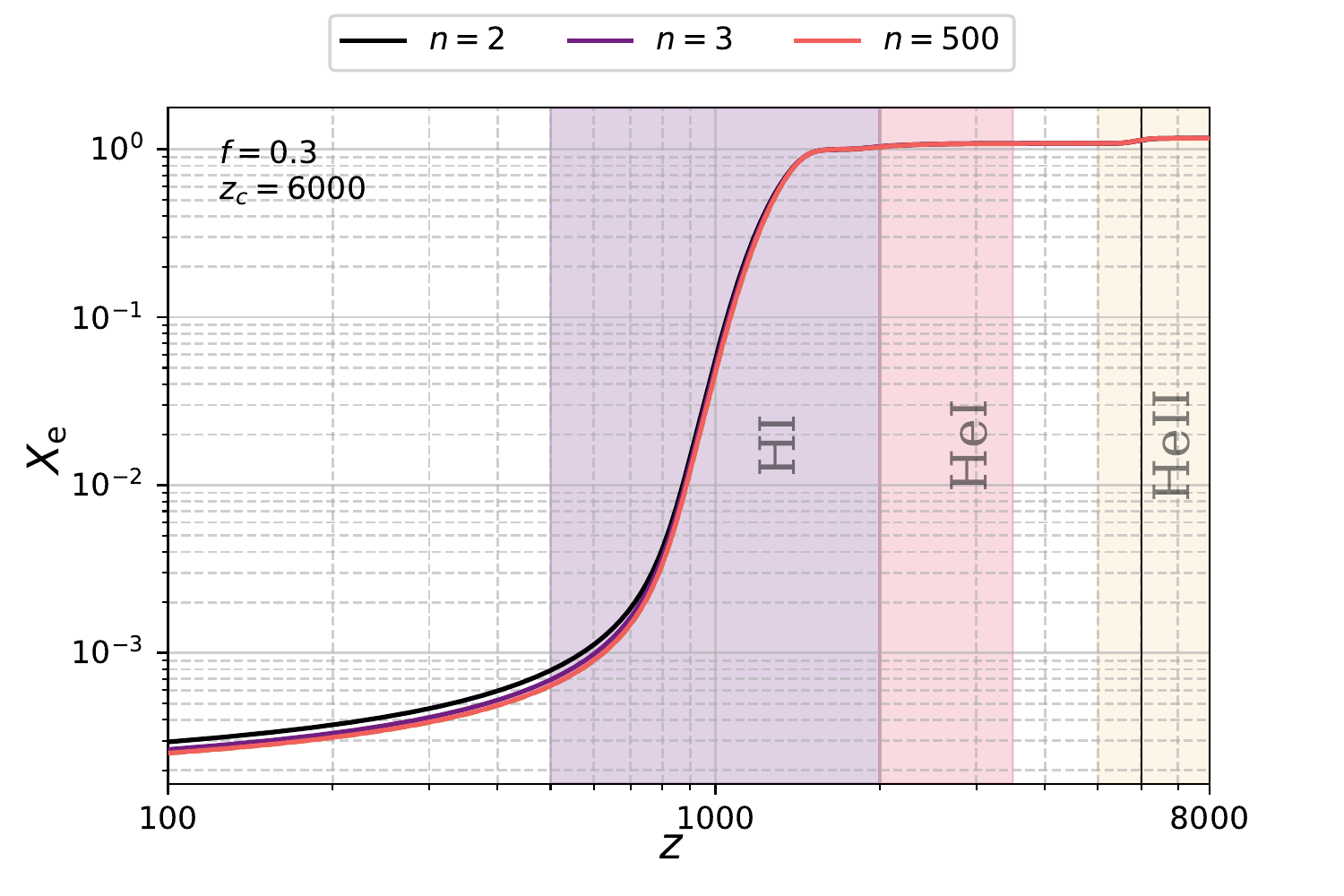}
    \\
    \caption{The ionisation history, $\xe$ for added early dark energy contributions§ with $\zc = 1500$ (\emph{top}) and $\zc = 6000$ (\emph{bottom}) compared to the \LCDM case (gold/dashed). The different order of potentials $n\in\{2,3,500\}$ is shown by the colour scheme \{black, purple, orange\}, respectively. The critical redshift, $z=\zc$ is also shown by a vertical black line.} 
    \label{fig:ede_xe}
\end{figure}

\begin{figure}
    \centering
    \includegraphics[width=\columnwidth]{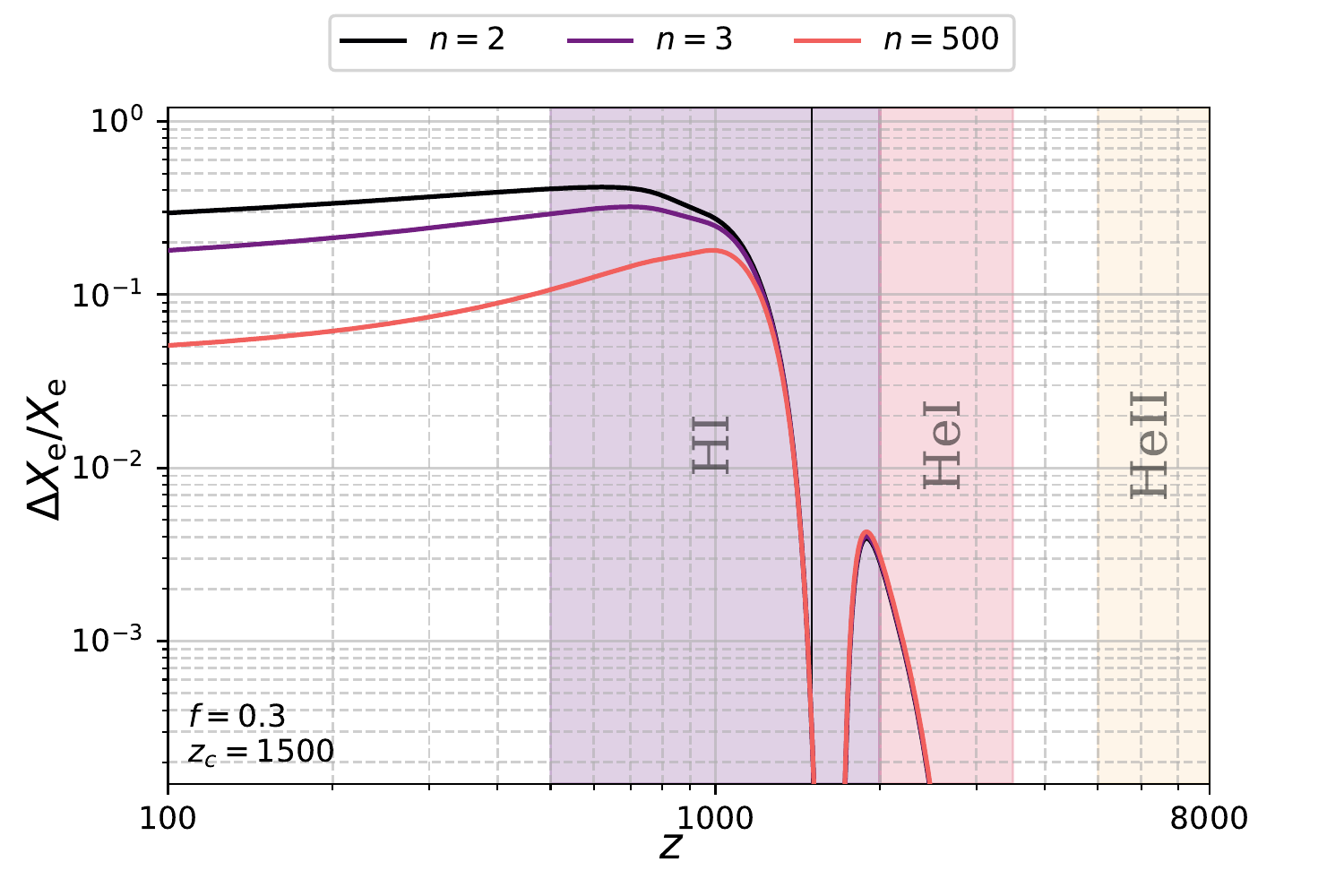}
    \includegraphics[width=\columnwidth]{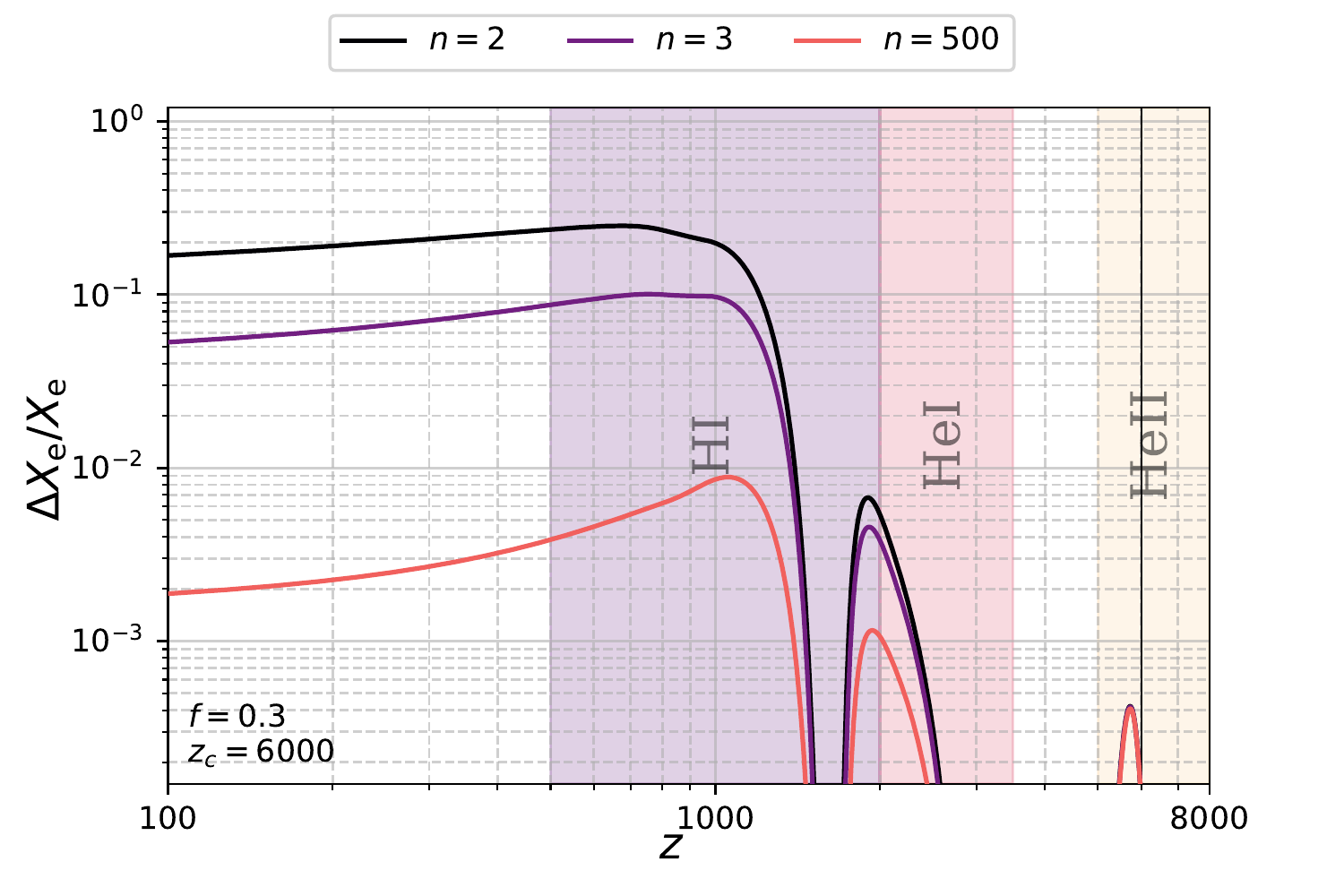}
    \\
    \caption{Variations in the ionization history arising from $\fede = 0.3$ and dynamical redshifts $\zc=1500$  (\emph{top}) and $\zc=6000$  (\emph{bottom}). For both cases, the different dilution rates for $n\in\{2,\,3,\,500\}$ are shown. Rough guidelines for the different recombination eras are also included.}
    \label{fig:ede_dxe}
\end{figure}

Here will we discuss the impact of an early dark energy contribution to the recombination lines so it is important to emphasise that we will only consider the effects on the background physics. In this case we do not consider the full changes to perturbations carried out in previous analyses, rather we consider the changes to the recombination process due to the altered expansion history shown in Fig.~\ref{fig:ede}. Furthermore, we ensure that flatness is conserved for these models (i.e., $\Omega_\Lambda\rightarrow \Omega_\Lambda + \Omega_{\phi,0}$). 
\changeJ{This condition should not directly affect the recombination lines, as $\Omega_\Lambda$ only becomes important at late times.}
The recombination calculations are carried out using {\tt CosmoRec} \citep{Chluba2010b} with the {\tt CosmoSpec} module \citep{Chluba2016CosmoSpec} to compute the CRR. This allows us to cleanly propagate all the effects on the hydrogen and helium recombination dynamics.

Current constraints on $\fede$ derived from CMB and large-scale structure measurements imply $\fede\simeq 0.05-0.1$ for critical redshifts $\zc\simeq 3000-4000$ [and fixed $n\simeq 3$] \citep{Hill2020EDE, Simon2022, Cruz2022}. To better illustrate the effects on the CRR, we will use larger values of $\fede\simeq 0.3-0.5$ and also widen the range of critical redshifts explored for varying values of $n$ as stated. The value of $\fede$ will mostly lead to an overall rescaling of the corresponding signals, while both $\zc$ and $n$ affect the shape of the distortion responses, as we illustrate below.

\subsection{Ionization history effects}\label{sec:ede_xe}
The modifications to the background dynamics propagate to the ionisation history. The changes to the Hubble rate $H(z)$ alter the total energy density of the Universe as a function of time and affect the mapping to redshift (see Fig.~\ref{fig:hz}). Consequently, recombination is delayed as can be seen from the ionisation history ($\xe$) variations shown in Fig.~\ref{fig:ede_xe}. These are compared to the case for standard \LCDM (gold/dashed). For all dilution rates, the variations in $\xe$ are larger for a lower critical redshift, $\zc = 1500$. Due to the smaller relative helium fractions \citep[and helium feedback processes at $z\simeq 2000$,][]{Chluba2009c, Chluba2012HeRec}, the relative changes in the ionisation history are more pronounced during hydrogen recombination, which is expected to lead to greater variations in the Thomson visibility function (since the Thomson visibility is larger around the last scattering epoch at $z\simeq 1100$). 

However the residuals of the ionisation history shown in Fig.~\ref{fig:ede_dxe} reveal the bigger picture for the earlier epochs of recombination. For both cases ($\zc = 1500$ and $\zc = 6000$), this corresponds to a positive residual in the ionisation history ($\dxe>0$). When the critical redshift (the time when the field becomes dynamical and the energy density dilutes) is shifted from the hydrogen recombination era ($\zc = 1500$) to the doubly-ionised helium ($\text{HeIII}\rightarrow\text{HeII}$) recombination era ($\zc = 6000$), modifications in the epochs of helium recombination become visible. Specifically, a noticeable variation around $z\simeq 6000$ arises in the free electron fraction, while the changes during hydrogen and neutral helium recombination show an increased sensitivity to the value of $n$. This highlights that the CRR can in principle be used as a probe of EDE. In particular for models with $\zc\gtrsim 5000$ this could nicely complement probes based on the CMB anisotropies alone, which already tightly constrain scenarios with $\zc\lesssim3000$ \citep{Simon2022}.


\subsection{Propagating changes to spectral distortions}\label{sec:ede_dist}
The changes to the ionization history that arise from early dark energy model extensions can be propagated into the deviations of the CRR using {\tt CosmoSpec}. In this section, we isolate some of the key features that have been modified in the CRR by considering an EDE species added into the expansion rate.

\subsubsection {Hydrogen recombination lines}
\label{sec:hEDE}

\begin{figure}
    \centering
    \includegraphics[width=\columnwidth]{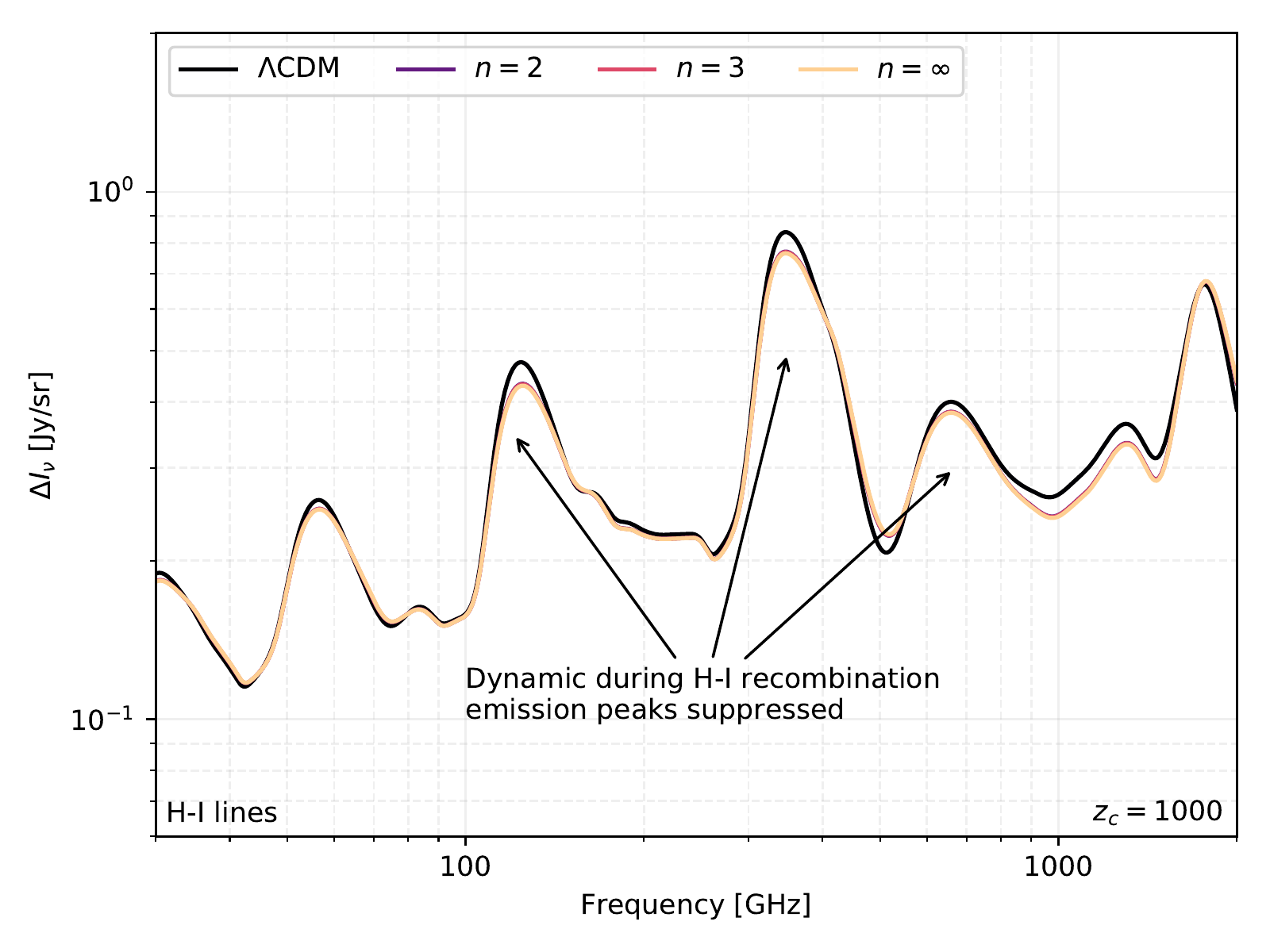}
    \\
    \caption{Hydrogen recombination lines for EDE models fixed with a dynamical time of $\zc = 1000$ for a variety of dilutions [$n=\{2,3,\infty\}$]. These are compared against the \LCDM benchmark (\emph{black}) over the frequency of interest $\nu \simeq 30-2000$ GHz.}
    \label{fig:hydrogenEDEzc1000}
\end{figure}

In Fig.~\ref{fig:hydrogenEDEzc1000}, the hydrogen recombination lines are shown for \LCDM (\emph{black}) and against the added EDE fluid with an `late' dynamical time of $\zc=1000$. For illustration, we vary the dilution rate ($\leftrightarrow n$) at fixed $\fede=0.5$.
There is very little movement in the line positions; however, the amplitudes of most spectral features is suppressed, including the Balmer-$\alpha$ line ($\nu\simeq120$ GHz) and Paschen-$\alpha$ line ($\nu\simeq350$ GHz). In addition, the lines are broadened due to the more extended duration of the recombination process. 
At $\nu \gtrsim 1800\,{\rm GHz}$, we can furthermore see an increased blue-wing of the HI Lyman-$\alpha$ line while the level of the two-photon continuum ($\nu \simeq 800-1500\,{\rm GHz}$) is reduced, indicating a delay in the recombination process.
Since the hydrogen line emission process occurs at $z\simeq 1400$, for the chosen example the variations in the lines appear invariant with the speed of dilution (affected by $n$) for this model.

\begin{figure}
    \centering
    \includegraphics[width=\columnwidth]{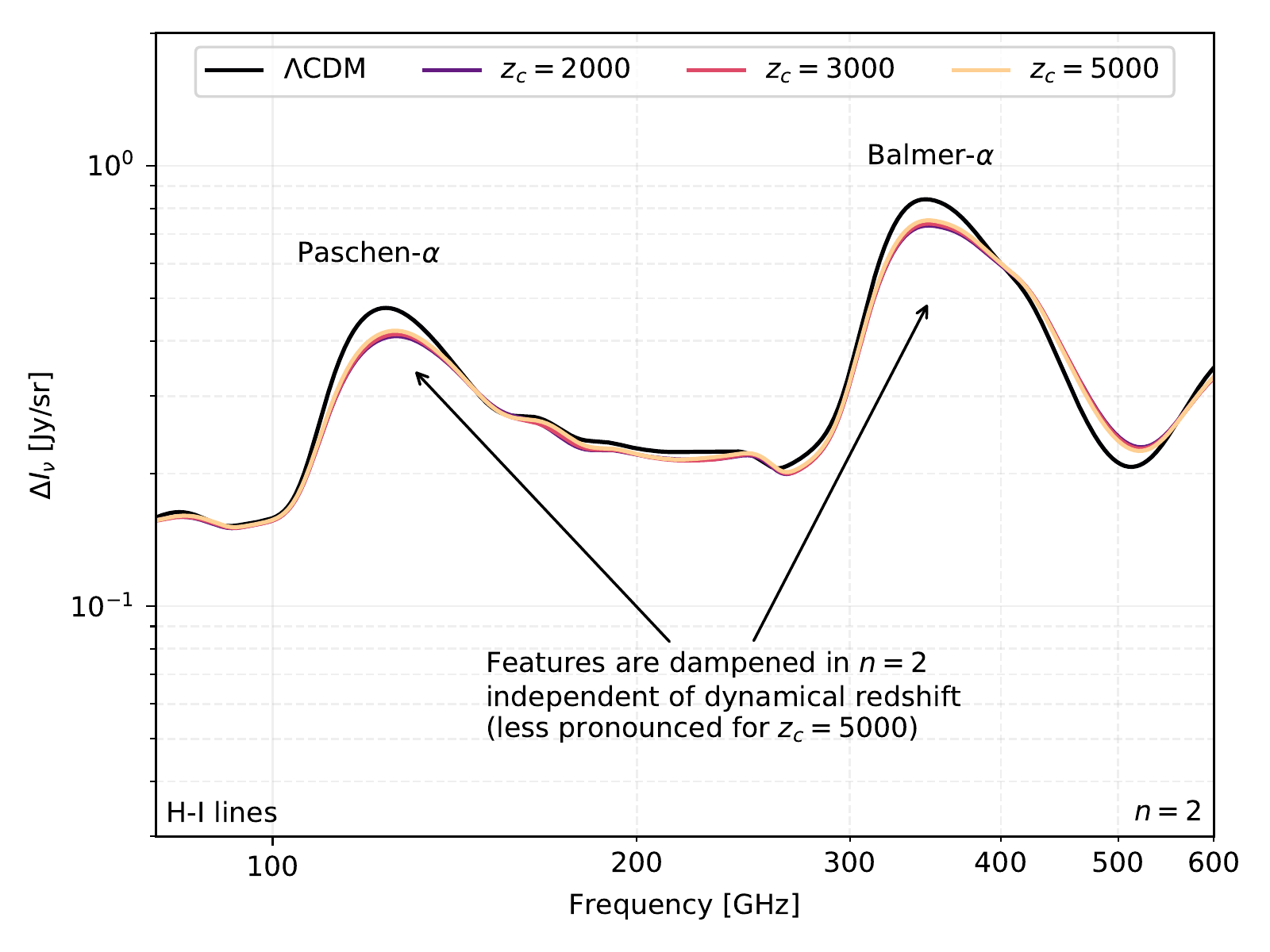}
    \\
    \caption{Example of effects on the Balmer-$\alpha$ (H$\alpha$) and Paschen-$\alpha$ lines caused by EDE modifications. Here the ULA field dilutes like cold-dark matter however is shown for a variety of dynamical redshifts, all showing suppression of the emission peaks.}
    \label{fig:hydrogenEDEN2peaks}
\end{figure}

In Fig.~\ref{fig:hydrogenEDEN2peaks}, we focus our attention on the Balmer-$\alpha$ and Paschen-$\alpha$ lines. 
Here, the ULA treatment of EDE is calculated with a variety of dynamical times and a radiation-like dilution ($n=2$). 
For $n=2$, we can see that even the changes of the CRR are mostly independent of the chosen dynamical time, though there is a very small ($\lsim 5\%$) effect for $\zc=5000$, which is marginally closer to the \LCDM case. %
The main conclusion from both Figs.~\ref{fig:hydrogenEDEzc1000} and~\ref{fig:hydrogenEDEN2peaks} is that the variations for slower dynamical EDE models closer to the surface of last scattering are qualitatively indifferent for hydrogen recombination lines. 

\begin{figure}
    \centering
    \includegraphics[width=\columnwidth]{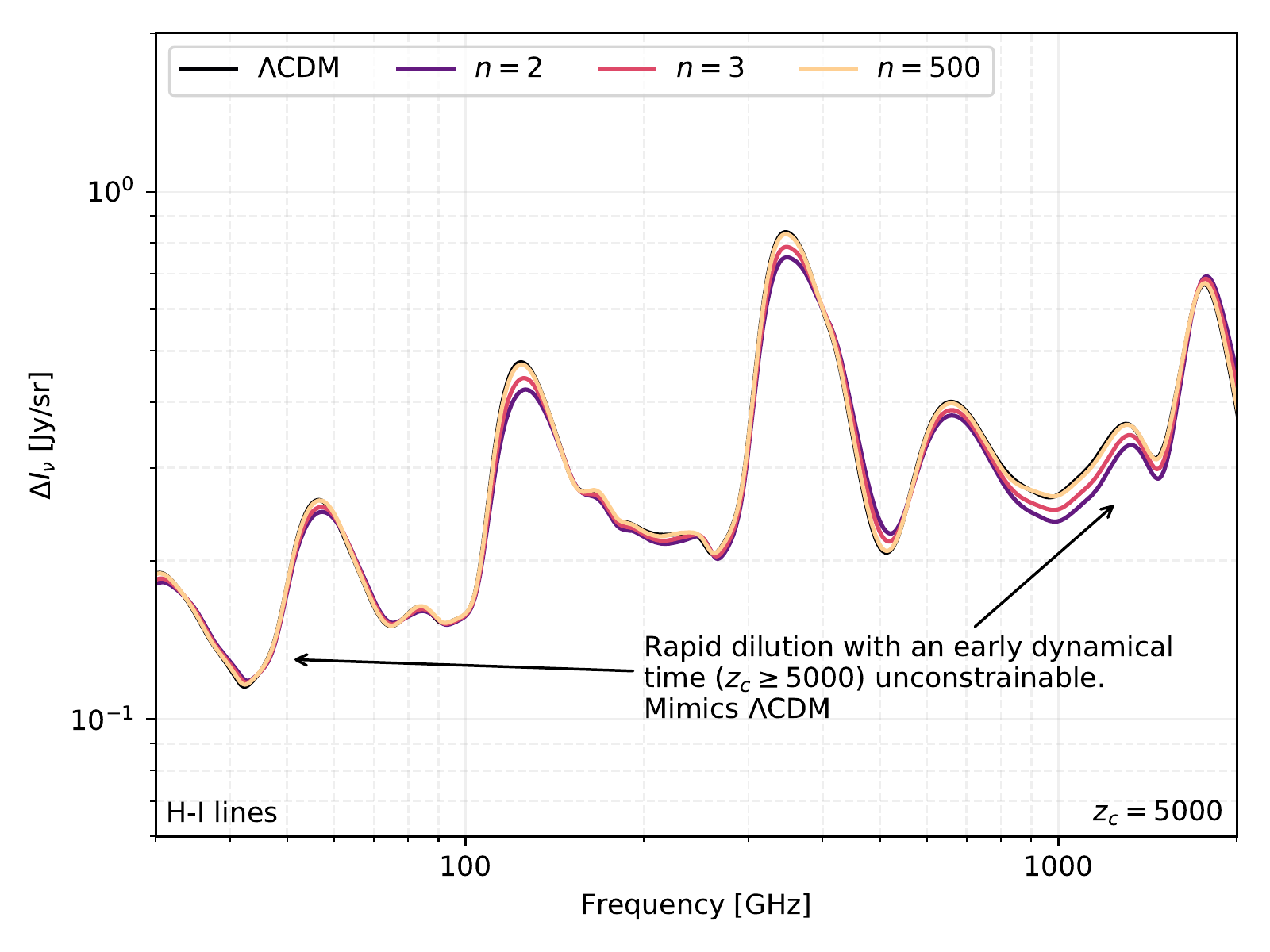}
    \\
    \caption{Hydrogen lines over the same frequency range as shown in Fig.~\ref{fig:hydrogenEDEzc1000}, however focussing on a dynamical time of $\zc = 5000$. The variations arising from the EDE models are much smaller in comparison to the model described by the aforementioned figure.}
    \label{fig:hydrogenN500weak}
\end{figure}

We present the contrary to this in Fig.~\ref{fig:hydrogenN500weak}, where we focus on a pre-recombination dynamical time, $\zc=6000$. For the different dilution models, the changes compared to \LCDM are heavily hindered, with the kinetic ULA example ($n\rightarrow\infty$; \emph{orange}) being almost identical to the \LCDM case. For a dynamical field that dilutes exponentially fast, the variations do not seed in the hydrogen lines at all. Since the ULA contributions to the energy density dilute away very quickly, the hydrogen recombination process is shielded from these modifications to the expansion rate. In comparison, and with reference to the species dilution curves in Fig.~\ref{fig:ede}; the denser EDE models such as radiation-like ($n=2$) and ultra-relativistic ($n=3$) tail off more slowly, since the dilution of the field is much harder. Consequently, the net impact on the expansion rate is larger and the interplay across the hydrogen recombination lines for an early dynamical time such as $\zc=5000$ becomes more apparent, as presented in Fig.~\ref{fig:hydrogenN500weak}. This shows that the CRR is sensitive to the dilution rate of EDE models with $\zc\gtrsim 3000$, as already anticipated from Fig.~\ref{fig:ede_dxe}.

\subsubsection{Helium recombination lines}\label{sec:heEDE}

\begin{figure}
    \centering
    \includegraphics[width=\linewidth]{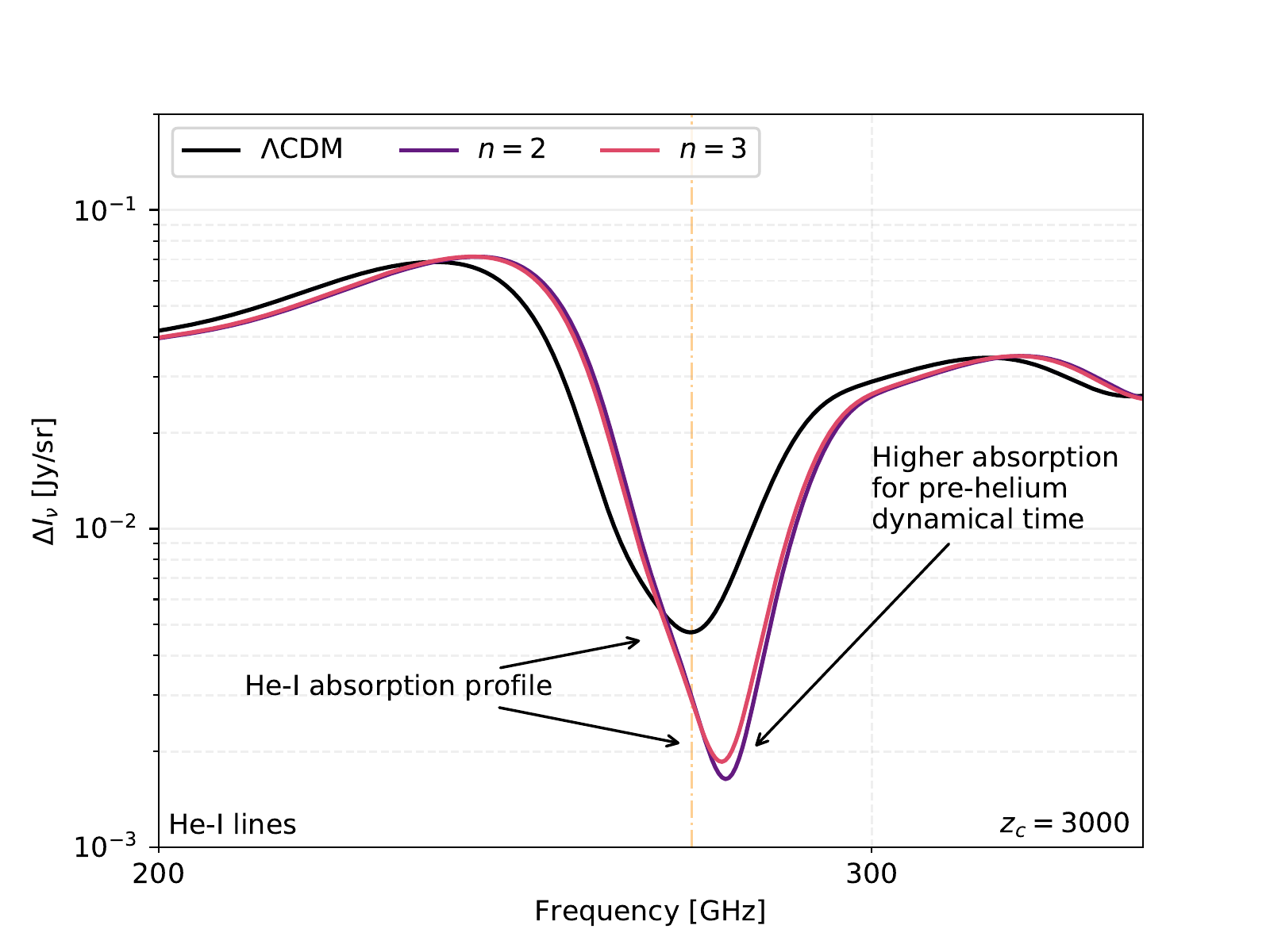}
    \caption{The absorption trough in the \changeJ{He-I radiation} at $\nu\simeq260$ GHz. The matter and radiation-like dilution models are shown for a fixed dynamical time associated from before He-I recombination ($\zc = 3000$).}
    \label{fig:heIabsorption}
\end{figure}

As emphasised in Sec.~\ref{sec:ede_xe}, the helium recombination lines will be more susceptible to changes in the expansion rate encroaching on earlier epochs ($z>2000$). However given the lower impact of helium recombination, both within the lines and the optical depth of CMB photons arising during the decoupling era, we expect this effect to be unique but smaller.
One of the most distinctive features in helium recombination is the backwash of photons arising from feedback processes between He-I and He-II. In Fig.~\ref{fig:heIabsorption}, the absorption trough for helium where $\nu\simeq270$ GHz is shown. By adding the diluting field that emulates EDE, the trough shifts to higher frequencies and dips to a weaker signal ($\Delta I_\nu\simeq0.002$ Jy/sr). Furthermore, the profile defining this `absorption' is much sharper for an EDE model. The earlier dynamical time starts to impact singly-ionized helium recombination (He-II$\rightarrow$He-I) as the Hubble flow accelerates before this epoch. Consequently, the energy rates involved are naturally dampened by the increased expansion making the effects from the absorption sharper than in the \LCDM case. 

\begin{figure}
    \centering
    \includegraphics[width=\linewidth]{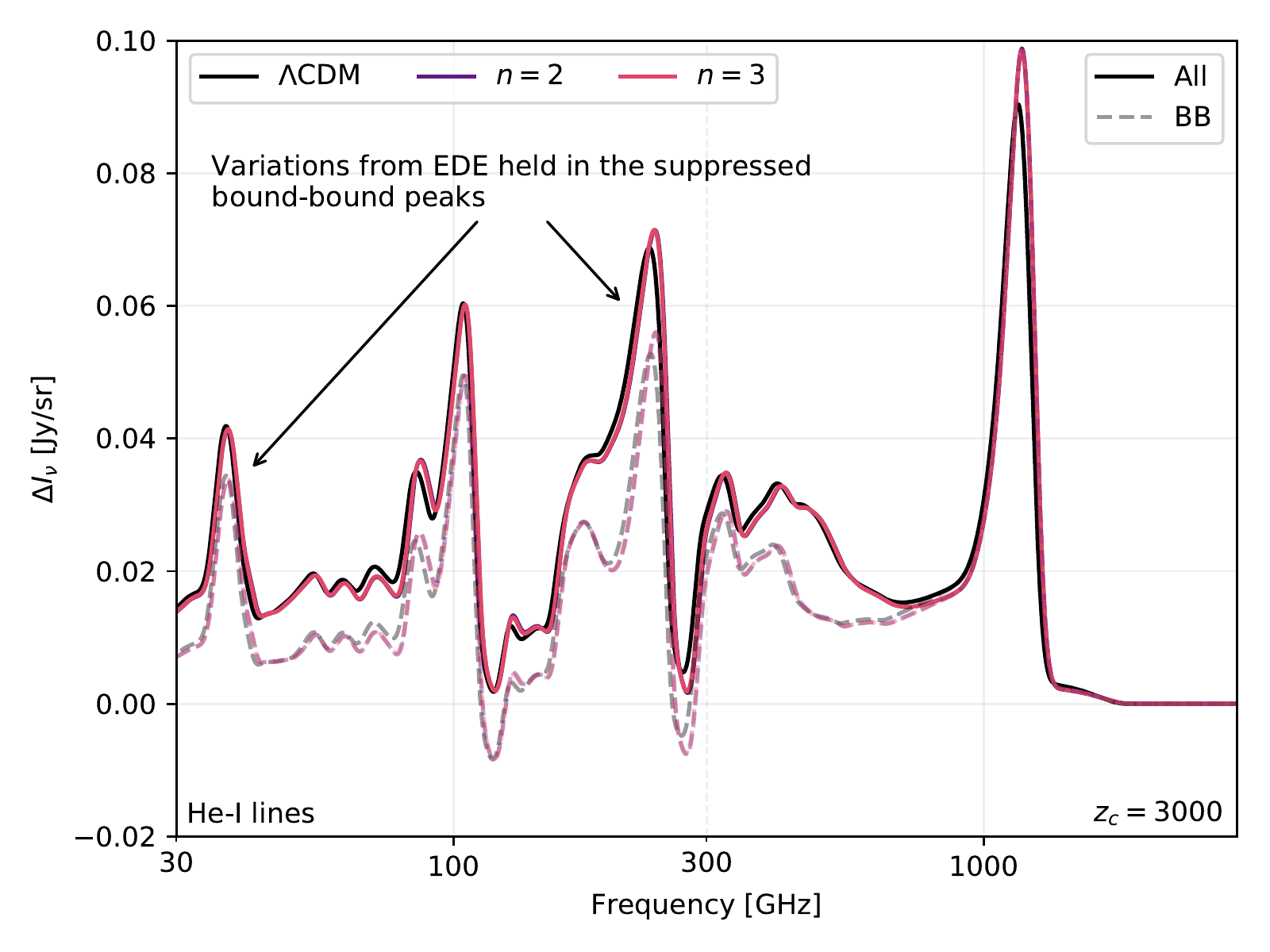}
    \caption{Helium-I recombination lines from Fig.~\ref{fig:heIabsorption} with wider frequencies to show the impacts on the \LCDM case and the bound-bound (BB) spectra when a similar EDE model is added (e.g., $\zc=3000$,$n=\{2,3\}$).}
    \label{fig:heIbb}
\end{figure}

A wider example of this absorption feature can be seen if we look at the full \changeJ{He-I} recombination spectra between the frequencies that are relevant for future space missions\footnote{In the  \emph{PIXIE} \citep{Kogut2011PIXIE} and \vtwenty \citep{Chluba2019Voyage} mission concepts, the frequency bands were defined for $\nu_{\rm min} = 30$ GHz and $\nu_{\rm max} = 3000$ GHz.}. This is highlighted in Fig.~\ref{fig:heIbb}, where we have also shown the impact of the full distortion against the case where bound-bound transitions are considered only (\emph{faded, dashed}). Removing the logarithmic $y$-axis, the rich structure of the helium emission and absorption features are clearer to see; however, the impact of the EDE changes does not reveal added information when we omit the bound-free absorption components in the distortions (free-free typically affects much lower frequencies). The contrast between amplification of the peaks and suppression of the troughs in the wider frequency information for helium also suggests that the expansion's impact on the net energy transitions is not trivial, due to the interlinked nature of He-I recombination to both the later hydrogen recombination epoch and the earlier doubly-ionised helium region.

\begin{figure}
    \centering
    \includegraphics[width=\linewidth]{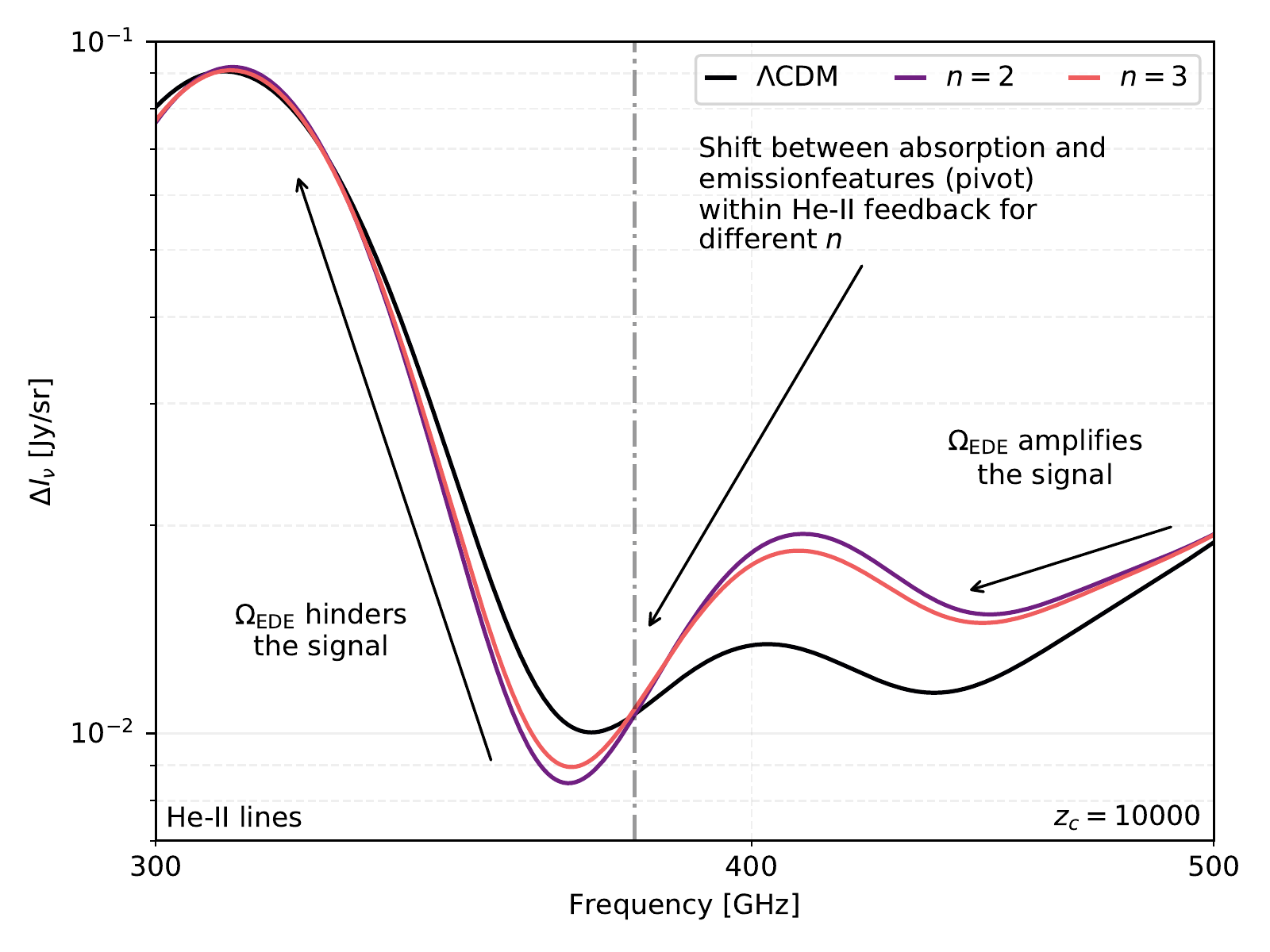}
    \caption{The effects on the He-II distortion arising from an EDE model with very early dynamical time ($\zc=10000$) for $300$ GHz $<\nu<500$ GHz. The vertical line at $\nu=378$ GHz acts as a pivot for the variations when EDE is introduced. This leads to amplification of the signal at $\nu>378$ GHz and suppression at $\nu<378$ GHz. There is subtle smaller pivoting in the accompanying helium resonance at $\nu\simeq320$ GHz.}
    \label{fig:heII_shifts}
\end{figure}

\begin{figure}
    \centering
    \includegraphics[width=\linewidth]{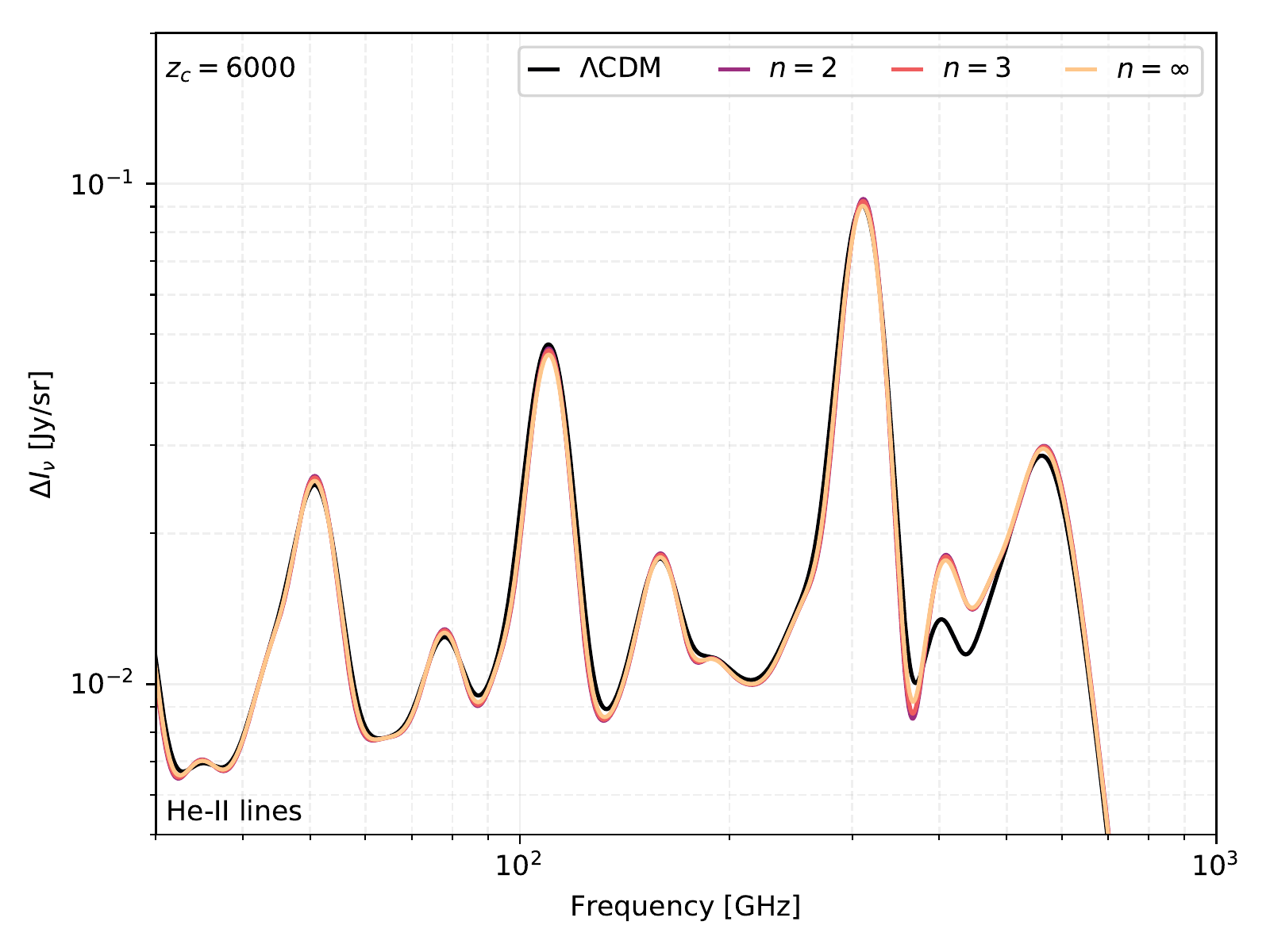}
    \caption{The full distortion arising from doubly-ionised helium recombination between the proposed \emph{PIXIE} bandwidths ($30$ GHz $<\nu<3000$ GHz). Here the different EDE models are shown with various dilution speeds as we have previously discussed ($n=\{2,3,500\}$) for a critical redshift $\zc=6000$.}
    \label{fig:heII_all}
\end{figure}

Finally, we focus solely on the doubly-ionised helium ($\text{HeIII}\rightarrow\text{HeII}$) region, which occurs significantly earlier ($z\simeq 6000$). In Fig.~\ref{fig:heII_shifts}, the absorption between the two helium species is shown ($300-500$ GHz) for \LCDM against two dilution examples. The line at $\nu\simeq380$ GHz represents a pivot for the added EDE models. At lower frequencies, this feature in helium is suppressed whereas it is amplified for higher frequencies. The full recombination spectrum for doubly-ionised helium is shown for context in Fig.~\ref{fig:heII_all}, however this was shown for $\zc=6000$ since it is more directly related to the doubly helium recombination era. The variations according to early dark energy all show distinct changes to the \LCDM CRR; however, the different dilution models do not seem to create appreciable changes in the spectra for the chosen value of $\zc$. Whether the field dilutes as radiation or ultra-radiation, the spectra look very similar. 
%

%
These changes arise because, as with previous cases, the dynamic changes for these models are harder to distinguish for earlier redshifts (i.e., during He-III$\rightarrow$He-II recombination). However, the knee-pivot discussed in Fig.~\ref{fig:heII_shifts} at $\nu\simeq400$ GHz is an isolated large change which can be associated with continuum processes. These continuum features \citep[discussed in more detail in][]{Switzer2007I, Jose2008} can be amplified if the electrons recombine into the bound states. This is more emphasised with states at higher energies, that are simultaneously closer to the ground state. There is also interplay with the fine structure lines in the helium atom \citep[see][for more details]{Chluba2012HeRec}. 
\begin{figure}
    \centering
    \includegraphics[width=\linewidth]{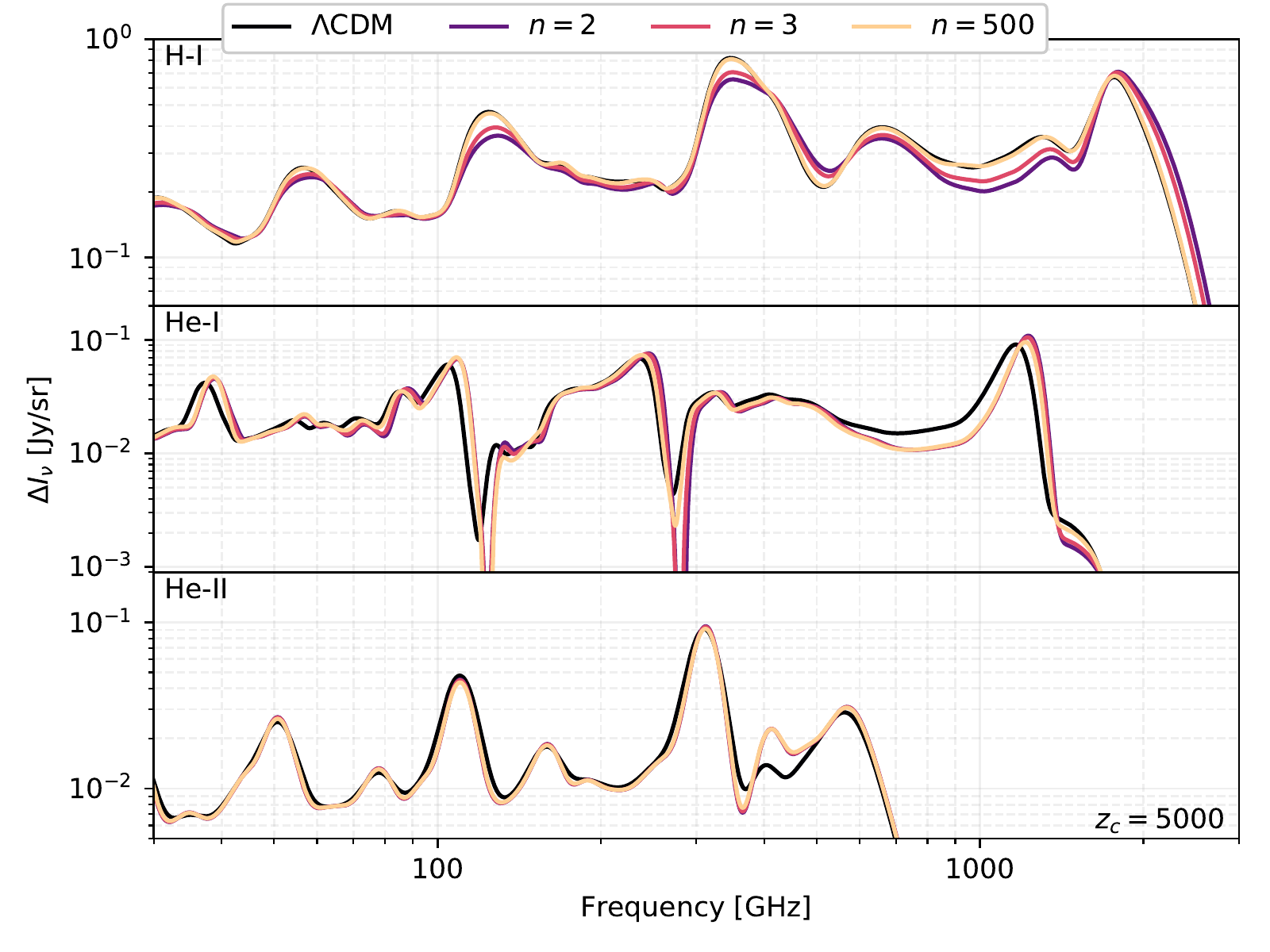}
    \caption{Comparison for an EDE model that has different slopes for $\zc=5000$ across the different atomic species. Hydrogen (\emph{top}) is compared against helium I and II (\emph{middle,bottom}) respectively. For better clarity, this has been plotted with $\fede = 0.8$.}
    \label{fig:h_all}
\end{figure}
\subsubsection{Combined effect from all the atomic species at $\zc=5000$}\label{sec:total5000}
To illustrate the total impact of various models when the early dark energy starts diluting right in the middle of the two helium recombination eras, we show the effects on the hydrogen and two helium energy spectra in Fig.~\ref{fig:h_all}. For the purposes of this figure, we have increased the relative amount of EDE to $\fede=0.8$. While the hydrogen lines are showing signs of spectral deviations according to the dilution slope, $n$; the helium atomic variants are showing this less prominently. There are difference tracers in the lower frequency ($\nu\simeq200$ GHz), particularly around the He-I absorption trough; however, the effects are less apparent in the helium plateau at $\nu\simeq800$ GHz and there is negligible spectral variance in the doubly-ionised helium lines.
This is anticipated, since the lever arm is shortest for the He-II spectrum and biggest for the H-I radiation, rendering the $n$-sensitivity largest for H-I.
These variations are hinting that the EDE mechanism can begin to act as a pivot: highlighting not only changes in different magnitudes across the epochs, but more importantly that the impact of the sloping parameter $n$ is non-negligible as you tune your critical redshift to earlier times. 

\subsection{Basic forecast for future spectrometers}\label{sec:forecastEDE}
In this section, we use a basic signal-noise ratio (SNR) test to validate the most responsive EDE models affecting the recombination lines. From modelling the various combinations of the dilution speed and critical redshift for the EDE theories (i.e., $n$ and $\zc$), we can quantify the models with the largest net change to the \LCDM reference using a basic signal-to-noise (diagonal Fisher matrix) testing case \citep[statistical use cases explained in][]{Sellentin2016,Bhandari2021}. Since the early dark energy fraction parameter ($\fede$) is the amplitude of the signal and is fairly model-independent (larger values of $\fede$ imply larger responses) then this is the parameter for which we model an effective signal-to-noise ratio to identify exceptional models.

\begin{figure}
    \centering
    \includegraphics[width=\linewidth]{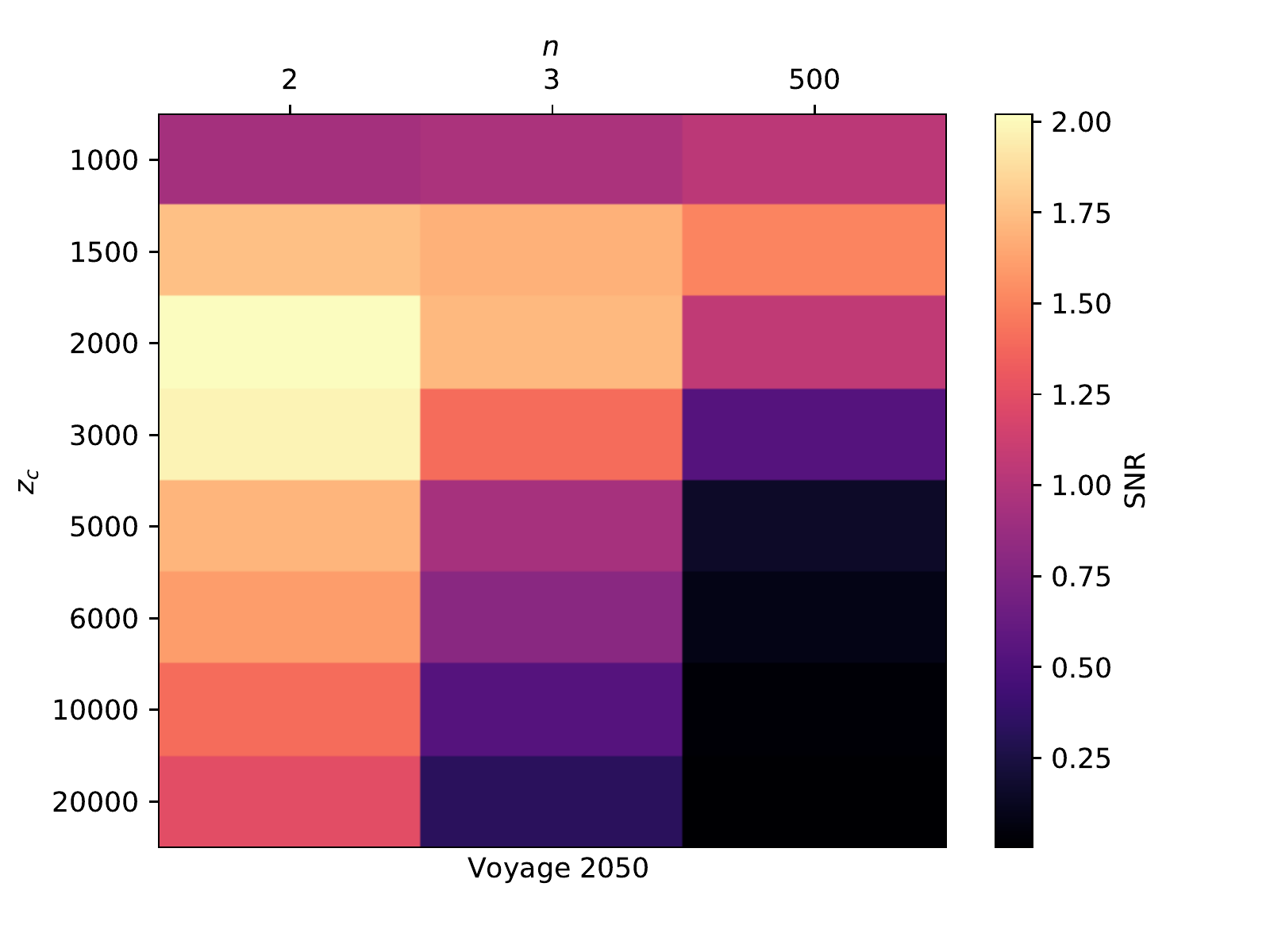}
    \caption{Signal-to-noise forecasts for \vtwenty for various EDE models defined by $n$ and $\zc$. The largest SNR values corresponds to the brightest colours in the colorbar. Similar SNR distributions can be found for \emph{SuperPIXIE} and \vtwentyp by multiplying the SNR by the appropriate change in sensitivity. For this particular study, $\fede = 0.05$. }
    \label{fig:edeMatrix}
\end{figure}

For this simple case study, we have isolated the frequency bins using the proposals for channels outlined for \emph{SuperPIXIE}, where we consider 3 instruments: low-frequency ($10$ GHz $<\nu<40$ GHz; $\Delta\nu = 2.4$ GHz), mid-frequency ($20$ GHz $<\nu<600$ GHz; $\Delta\nu = 19.2$ GHz) and high-frequency ($400$ GHz $<\nu<6000$ GHz; $\Delta\nu = 57.6$ GHz). The full details of this spectrometer setup can be found in many forecast proposal papers \citep{Kogut2011PIXIE,Abitbol2017, Chluba2019Voyage, Hart2020c}. The signal-to-noise is isolated in each bin and then the summation RMS value is taken for every given model. After evaluating each of the known dilution models [$n = \{2,3,\infty\}$] for a wide range of critical redshifts [$\zc = \{1000,1500,2000,3000,5000,6000,10000,20000\}$], the full response of each model can be seen in Fig.~\ref{fig:edeMatrix}. Here the specifications for a spectrometer following the \vtwenty outline \citep{Hart2020c} have been used to calculate an effective SNR. Note that this process omits correlated parameters but we can start to implement a basic Fisher matrix analysis and see how the parameter dependencies vary for different EDE models. 

The SNR values for these models have been visually summarised in Fig.~\ref{fig:edeMatrix}. The top five models have been quantitatively presented as a list in Table~\ref{tab:snrEDE}. As initially predicted from the large hydrogen variations in Sec.~\ref{sec:ede_dist}, the strongest models for a \vtwenty like mission have a slope consistent with a radiation-like dilution ($n=2$). Constraining the EDE model requires a \emph{`sweet spot'} solution where the field dilutes quickly enough that the net change in dynamics across recombination is large, while not requiring such a high initial value that may affect the initial conditions. These initial conditions will ultimately affect the background as recombination lines begin to form and the non-thermal processes between electrons and photons emerge. More specifically, the variations favour a dilution that begins in the HeI recombination era ($z\simeq2000,3000$) since this provides the largest amount of variation within the recombination radiation for the smallest additional factor of EDE. The most constraining detectable model according to this comparison is the model is a radiation-like dilution with $\zc = 2000$ with an SNR of $2.02$ for \vtwenty, as shown in Table~\ref{tab:snrEDE} ($M_A$). Note that the results shown for the \spixie and \vtwentyp configurations are multiplied by constant factors of $0.2$ and $10$ respectively, as their noise curves are defined.
Surprisingly, for radiation-like dilution, a redshift switch of $\zc=5000$ has shown similar SNR in Table~\ref{tab:snrEDE} (Model $M_E$), suggesting that some of the pre-HeI recombination models can survive. Large variations from models like these will still appear during hydrogen recombination; however, as shown in Fig.~\ref{fig:edeMatrix}, the sharper dilution models quickly deteriorate to low SNRs once the EDE phase-transition redshift is pushed to earlier times. 

In summary, we can see that for radiation-like dilution ($n=2$), a wide range of critical redshifts may be probed, even reaching deep into the primordial universe at $z\gtrsim 10^4$. For steeper dilution, sensitivity to $\zc\gtrsim 6000$ quickly drops, and lower values of $\zc$ give preferable responses.

\subsection{Using Fisher matrix analysis to estimate the errors}
\label{sec:fisherEDE}
From the leading order estimates on the strength of the EDE amplitude parameters, we can compute more realistic responses between the \LCDM parameters and $\fede$ for various models. 
In this section, we obtain Fisher forecasts for the 3 highest SNR models found in Table.~\ref{tab:snrEDE}. We discuss the correlations from model to model and refer to the main features of the spectral distortion variations highlighted in Sec.~\ref{sec:ede_dist}.\footnote{It is also important to note that we will revisit this methodology in Sec.~\ref{sec:vfc} where we apply a simpler version of the Fisher forecast to constraining fundamental constants with the recombination lines.}
One of the simplest statistical measures that we can use to test for parameter correlations is the Fisher information matrix that defines the covariances at the peak of the likelihood. The Fisher matrix is defined in many ways however for spectral distortions, the matrix is defined by,
\begin{equation}
    F_{ij} = \sum _{\nu \nu'}\frac{\partial \Delta I_{\nu}}{\partial p_i}\; {\bf{\Sigma}}^{-1}_{\nu \nu'} \; \frac{\partial \Delta I_{\nu'}}{\partial p_j},
\end{equation}
where $\{p_{i},p_{j}\}$ are the parameters in the correlation study that correspond to the Fisher matrix element $F_{ij}$. Here $\Sigma_{\nu\nu'}$ is the covariance matrix for the $\nu\times\nu'$ frequency bands. In this analysis, our covariance matrix is going to be made from the total signal coming from the fiducial $\Delta I_\nu$ spectra and the noise spectra discussed in Sec.~\ref{sec:forecastEDE}. 
Note that this is the same formalism of the Fisher matrix that was used in the previous paper constraining cosmological parameters with the recombination lines \citep{Hart2020c}. In Sec.~\ref{sec:resultsF}, we will present the covariances for the 3 most constrainable models (shown in \,Table.~\ref{tab:snrEDE}).
\changeL{The EDE amplitude $\fede$ is bound by a hard-prior where $\fede>0$. When sampling the Fisher matrices required for Fig.~\ref{fig:edeContours2}, we have not considered the physical limits such as those imposed by scalar parameters. Hard priors will be included in evaluating the posterior, designed for a full MCMC analysis. The methodology assumes perfectly Gaussian likelihoods and does not appropriately evaluate hard-priors for non-negative parameters (such as $\fede$ and $\Neff$). The errors calculated in this forecast are designed to be order-magnitude estimates for comparison and assessment against other probes.}

\begin{table}
    \centering
    \begin{tabular}{c|c|c|c|c|c}
        \hline\hline
        Model & $n$ & $\zc$ & SNR  & SNR  & SNR \\
        & & & (\emph{SuperPIXIE}) & (\emph{V2050}) & (\emph{V2050+})\\
        \hline
        $M_A$ & $2$ & $2000$ & $0.40$ & $2.02$ & $20.20$ \\
        $M_B$ & $2$ & $3000$ & $0.40$ & $1.98$ & $19.78$ \\
        $M_C$ & $2$ & $1500$ & $0.35$ & $1.76$ & $17.59$ \\
        $M_D$ & $3$ & $2000$ & $0.34$ & $1.72$ & $17.22$ \\
        $M_E$ & $2$ & $5000$ & $0.34$ & $1.70$ & $17.06$ \\
        \hline
    \end{tabular}
    \caption{The signal-to-noise ratios for the 5 most constrained models of EDE using the recombination lines (referred to as $M_i$), for 3 different model configurations. Note that \vtwentyp (\emph{V2050+}) is 10 times more sensitive than \vtwenty (\emph{V2050}). Here the dilution index $n$ and the dynamical redshift $\zc$ of each of these models are given along with the SNR (signal-noise ratio) value. \changeJ{We used $f_{\rm EDE} = 0.05$ as reference value.}}
    \label{tab:snrEDE}
\end{table}

\subsubsection{Contour results for the Fisher matrix}\label{sec:resultsF}
\begin{figure}
    \centering
    \includegraphics[width=\linewidth]{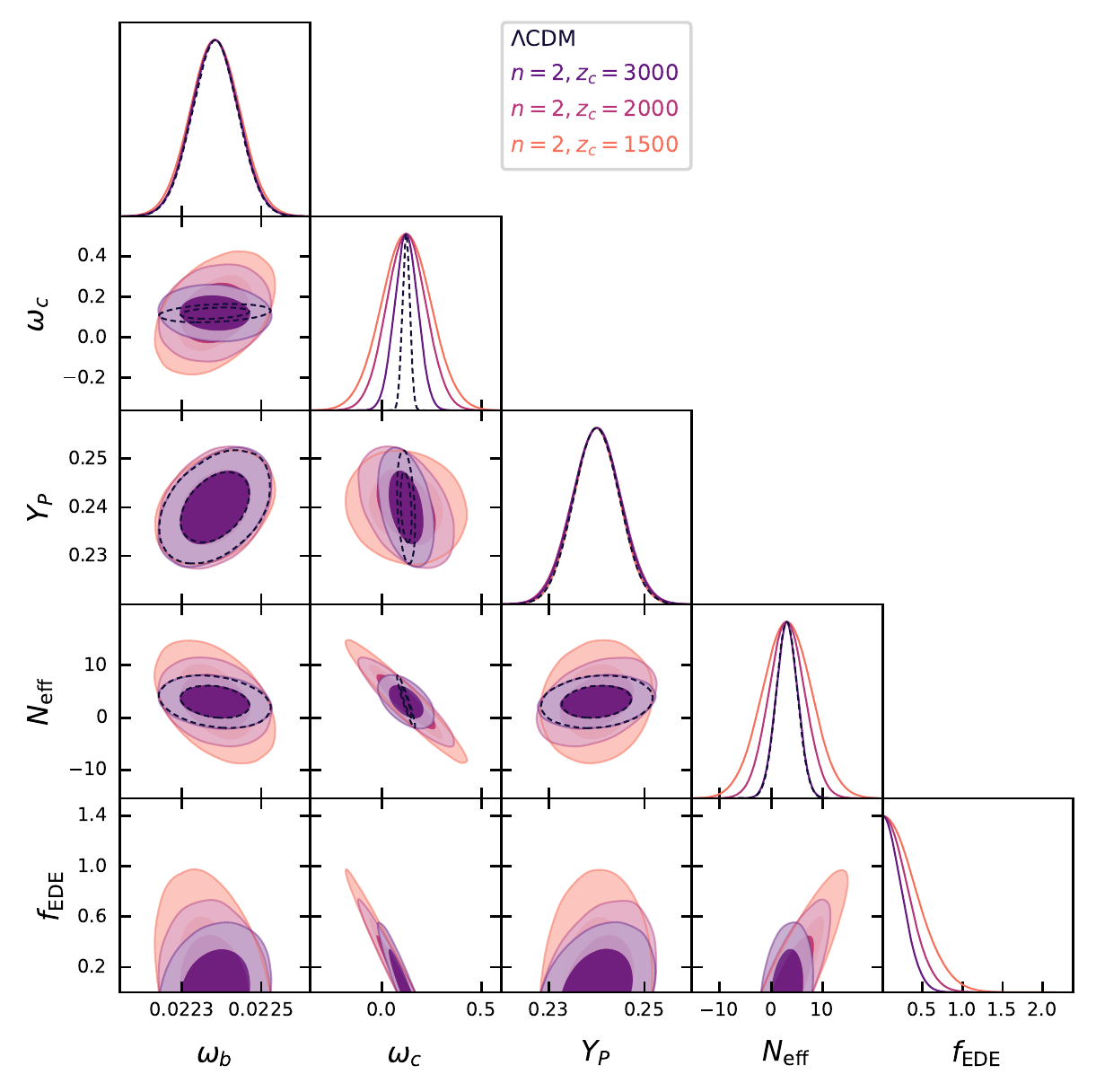}
    \\
    \caption{Estimated probability contours with a {\it Voyage 2050+} style mission for the EDE amplitude at $z=\zc$, $f_{\rm EDE}$. The fiducial values for the standard parameters arise from \planck and the fiducial value for the EDE amplitude is set to 0. The different colours refer to the three most promising models, whereas the black line indicates \LCDM.}
    \label{fig:edeContours2}
    \vspace{-3mm}
\end{figure}

Results from the Fisher matrix can tell us about the underlying parameter degeneracies in this EDE model. Using the Fisher matrix with a generation of random Gaussian samples, one can visualise the degeneracies using the same contours as in MCMC\footnote{The generation of Gaussian samples and subsequent plots were done with the {\tt GetDist} package \citep{GetDist}.}. The results for the 3 most promising models described in Sec.~\ref{sec:forecastEDE} are shown in Fig.~\ref{fig:edeContours2}. Note that the EDE amplitude parameter has been set to $f_{\rm EDE} = 0.0$ as the fiducial pivot point for the calculation. The stability of the step size for the Fisher matrix treatment was akin to the previous paper, as well as the diagonal Gaussian likelihood approach to the Fisher matrix \citep{Hart2020c}. 

All three models in Fig.~\ref{fig:edeContours2} have some degeneracies with $\omc$ and $\Neff$. Interestingly, the higher values of $\zc$ in this particular configuration get gradually more constrained. The contours for the $\zc=3000$ model have the smallest contours where the $1\sigma$ limit $\sigma_f\approx0.3$ assuming {\it Voyage 2050+}.
The marginalised errors attained with \planck for similar models are $\simeq 4$ times smaller \citep{Poulin2018, Hill2020EDE}. Not only do the contours for $n=2$ models show signs of deformation which can unknowingly bloat contours in simplistic analyses such as the Fisher presented here; it was shown in our previous paper that the expansion rate quantities would require a further $\simeq 50$ times greater sensitivity than \vtwenty to get the desired parameter constraints that would complement \planck. For these particular models, with a $50 \times$\vtwenty sensitivity, we could hope to constrain $f_{\rm EDE}\lsim0.05$ using only a CMB spectrometer. However this also neglects the involvement of foregrounds, albeit their broader spectral shape mitigates how badly they affect cosmological parameter constraints \citep[more details in][]{Hart2020c}.
Comparatively, for a \vtwentyp style mission, the model that begins to dilute during the peak of hydrogen recombination ($\zc = 1500$) is only at $f_{\rm EDE} \lsim 1$. This becomes the case as the contributions correlate directly with a matter-dilution effect on the CRR which primarily dominates hydrogen recombination: the fractional cold dark matter density today $\omega_c$. 

\changeJ{We note that at this stage we had no access to MCMC chains relating to the analysis of EDE models with \Planck. We therefore could not explore robustly how the addition of external data sets could help breaking parameter degeneracies. For the standard CRR analysis, we saw this to yield significant gains for $\Neff$ and $\Yp$ \citep{Hart2020c}. Below we will illustrate how in the case of VFCs the addition of \Planck priors indeed significantly improves matters. In a similar vein, we expect significant gains for EDE models when combining CRR measurements with external data set; however, a more detailed assessment is beyond the scope of this work.}

\vspace{-3mm}
\section{Varying fundamental constants}\label{sec:vfc}
One of the promising extensions to the standard \LCDM paradigm is the addition of variations to fundamental constants as we have alluded to in Sec.~\ref{sec:intro}. Specifically, recombination physics relies on the couplings between charged particles (electrons) and photons; therefore, the main constants that will affect this interaction are the fine structure constant ($\aEM$) and the effective electron mass ($\me$). In this section, we will recap the effect to the  recombination process arising from VFCs and succinctly discuss the updates of {\tt CosmoSpec} since the last analysis of the CRR. These updates proved vital for the accurate calculation of fine structure variations within the recombination epoch. Finally, we present the resulting spectral distortion changes from these parameters and use those in a Fisher forecast akin to Sec.~\ref{sec:fisherEDE} to get some zeroth level detectability estimates.

\subsection{Recap of VFCs in recombination}
\label{sec:recap_vfc}
Fundamental constants have a wide impact on various parts of cosmology, however the physics of recombination can be directly traced by variations in two constants: the fine structure constant $\aEM$ and the effective electron mass $\me$. These both affect many aspects of the atomic physics picture that directly impact the evolution of the free electrons during recombination. For reference, we present the summary table for state variables affected by these constants \citep{Scoccola2009,Hart2017}: 
\begin{align}
\label{eq:vfcScale}
\begin{split}
\sigma_{\rm T} \propto \alpha_{\rm EM}^2 m_{\rm e}^{-2} 
\qquad A_{2\gamma} &\propto \alpha_{\rm EM}^8 m_{\rm e} 
\qquad P_{\rm S} A_{1\gamma} \propto \alpha_{\rm EM}^{6}m_{\rm e}^{3} 
\\
\alpha_{\rm rec} \propto \alpha_{\rm EM}^2 m_{\rm e}^{-2} 
\qquad \beta_{\rm phot} &\propto \alpha_{\rm EM}^5 m_{\rm e} 
\qquad T_{\rm eff} \propto \alpha_{\rm EM}^{-2}m_{\rm e}^{-1}.
\end{split}
\end{align} 
We can extend this to the scalings to effective rate coefficients that are very important for full calculations to recombination as,
\begin{align}
\mathcal{A}_i\left(T_\gamma, T_{\rm e}\right) &\rightarrow \aEM^2\me^{-1}\,\mathcal{A}_i\left(\aEM^{-2}\me^{-1}\,T_\gamma, \aEM^{-2}\me^{-1}\,T_{\rm e}\right), \\
\mathcal{B}_i\left(T_\gamma, T_{\rm e}\right) &\rightarrow\aEM^5\me\,\mathcal{B}_i\left(\aEM^{-2}\me^{-1}\,T_\gamma, \aEM^{-2}\me^{-1}\,T_{\rm e}\right), \\
\mathcal{R}_{ij}\left(T_\gamma\right) &\rightarrow \aEM^5\me\,\mathcal{R}_{ij}\left(\aEM^{-2}\me^{-1}\,T_\gamma\right).
\end{align}
Here the effective recombination and photoionisation rates are $\mathcal{A}_i$ and $\mathcal{B}_i$ respectively, whereas $\mathcal{R}_{ij}$ represents the transitions between excited states \citep{Yacine2010}. For the spectral conductances \citep{Yacine2013RecSpec, Chluba2016CosmoSpec} that are required to calculate the CRR, the coefficients $\mathcal{G}_{n' n}^{X}$ are scaled by the same factor as the transition rates $\mathcal{R}_{ij}$. It is important to note that for these variations, we have not assumed a particular model and therefore assume no knowledge of an external field that could potentially manipulate the underlying background cosmology. This in turn means that we have not considered any modifications to the Hubble flow $H(z)$ arising from such fields. \footnote{This is a potential direction for more complex fundamental constant variations and may even couple to quintessence-like fields that resemble the EDE discussed in Sec.~\ref{sec:ede} \citep{Calabrese2011}.}

\vspace{-3mm}
\subsection{Modifications to {\tt CosmoSpec}}
\label{sec:mods_CosmoSpec}

\subsubsection{Rescaling effective conductances}
\label{sec:Gnu}
To include the effect of varying fundamental constants (VFCs) on the CRR, we follow the description presented in Appendix~B of \citet{Chluba2016CosmoSpec} and revisited in Sec.~\ref{sec:recap_vfc}. A few important differences with respect to the original version of {\tt CosmoSpec} are:
\begin{itemize}
    \item the \ion{He}{II} spectrum is now computed using rescaled conductances of \ion{H}{I}. This reduces the storage of data and we confirmed the results carefully by direct computation.
    \item the redshift range over which the conductances are tabulated was extended, as VFCs can allow recombination to occur at higher and lower temperatures than in the standard scenario.
    \item the effect of electron scattering is included for modified scattering cross section, with rescaled $y$-parameter obtained as $y'\propto \sigT/\me \simeq (\aEM'/\aEM)^2\,(\me'/\me)^{-3}\,y$ and updated recombination history.
    \item similarly the effect of free-free absorption is modelled with the optical depth scaled as $\tau_{\rm ff}'\simeq (\aEM'/\aEM)^3\,(\me'/\me)^{-1.5}\,\tau_{\rm ff}$ and modified recombination history.
\end{itemize}
Aside from these rather straightforward modifications we also improved the analytic treatment of photon escape from the main \ion{He}{I} resonances, as explained next.

\subsubsection{Treatment of \ion{H}{I} absorption during \ion{He}{I} recombination}
\label{sec:DPesc}
One of the important corrections to the helium recombination history at $z\simeq 1700-2000$ is the effect of neutral hydrogen continuum absorption, which leads to a significant acceleration of the recombination process \citep{Kholupenko2007, Switzer2007I, Jose2008}. To approximately model this process, one can compute the correction to the Sobolev escape probabilities of the main singlet and triplet resonances as \citep[see Appendix~B  of][]{Jose2008}:
\begin{align}
  \label{eq:DPesc_chi_1D}
  \Delta P^{1\rm D}_{\rm esc}
  &\approx\!\!\int_0^1 \!\!\!\id\chi\!
  \left\{1-\expf{-\tauS(1-\chi)}-\kappa(\chi)
  \left[1-\expf{-[\tauS+\tauct(\chi)](1-\chi)}\right]\right\},
\end{align}
where $\chi=\int_{-\infty}^\xD \phi(y, a)\id y$ is the integral over the Voigt-profile of the resonance, $\phi(\xD, a)$, with Voigt-parameter $a$; $\xD$ is the distance from line center at frequency $\nu_0$ expressed in Doppler-widths; $\tauS$ is the Sobolev optical depth of the line. We furthermore defined the \ion{H}{I} continuum opacity variables
\bsub
\begin{align}
\tauct(\chi)&=\frac{c\, N^{\rm H}_{\rm 1s} \sigma^{\rm H}_{\rm 1s}(\nu)}{H}
\,\frac{\Delta\nu_{\rm D}}{\nu}\,\frac{1}{\phi(\xD, a)},
\\[1mm]
\kappa(\chi)&=\frac{\tau_{\rm S}}{\tauS+\tauct(\chi)},
\end{align}
\esub
where one should think of $\nu$ and $\xD$ as functions of $\chi$. The Doppler width of the line is given by 
\begin{align}
\frac{\Delta\nu_{\rm D}}{\nu_0}&=\sqrt{\frac{2 k\Te}{\mHe c^2}}
\approx
\pot{1.7}{-5} \left[\frac{(1+z)}{2500}\right]^{1/2},
\end{align}
and directly depends on the electron temperature, with the typical value given for $\Te=\TCMB$.

It was shown that Eq.~\eqref{eq:DPesc_chi_1D} provides a good first approximation to the escape probability corrections of the \ion{He}{I} singlet and triplet resonances. This can be used to model the main effect of \ion{H}{I} continuum absorption on the recombination process and further corrections can then be added using the radiative transfer module of {\tt CosmoRec/CosmoSpec} \citep{Chluba2012HeRec}. 

Instead of computing the integral in Eq.~\eqref{eq:DPesc_chi_1D} repeatedly, in {\tt CosmoRec/CosmoSpec} the escape probability correction is pre-tabulated as a function of $\tauS$, $\etac=c\, N^{\rm H}_{\rm 1s} \sigma^{\rm H}_{\rm 1s}(\nu_0)/H$ and $\Te$. This eases the computation for the standard recombination history. However, to include the effect of varying fundamental constants, the strategy has to be slightly changed. The parameter $\tauS$ can be used as before, as it is computed internally for the rescaled atomic properties. The other dependencies on atomic properties and the electron temperature enter through the mapping of $\xD=\xD(\chi, a)$, which is a function of the Voigt-parameter $a$, and also $\nu/\nu_0=1+\xD(\chi, a)\,\Delta\nu_{\rm D}/\nu_0\equiv f(\chi, a, \Te)$. Overall, this suggests that $\Delta P^{1\rm D}_{\rm esc}=\Delta P^{1\rm D}_{\rm esc}(\tauS, \etac, \Te, a)$. 
By writing
\begin{align}
\tauct
&=
\etac\,
\frac{\Delta\nu_{\rm D}}{\nu_0}
\frac{1}{\phi(\chi, a)}\,\frac{\nu_0\, \sigma^{\rm H}_{\rm 1s}(\nu)}{\nu \, \sigma^{\rm H}_{\rm 1s}(\nu_0)}
\end{align}
we can realize that the explicit dependence of $\tauct$ on $\Delta\nu_{\rm D}/\nu_0$ can be captured by using $\etact=\etac\,\Delta\nu_{\rm D}/\nu_0$ as a new parameter for tabulation. Since in the optically-thick regime most of the contributions to the escape probability come from around $\nu\simeq \nu_0$, this in fact means that the explicit dependence on $\nu/\nu_0$ can be omitted. Consequently, a 3D table in $\tauS$, $\etact$ and $a$ should provide a good representation for the main scaling of $\Delta P^{1\rm D}_{\rm esc}(\tauS, \etac, \Te, a)\approx \Delta P^{1\rm D}_{\rm esc}(\tauS, \etact, a)$.

In practice, we can simply go back to the previous tabulation scheme but keep the electron temperature instead of the Voigt-parameter $a$. Since $a=A_{21}/[4\pi \Delta\nu_{\rm D}]\propto A_{21}/[\nu_0
\sqrt{\Te}]$, we have
\begin{align}
a'(\Te)=a(\Te) \, \frac{A'_{21}}{A_{21}}\frac{\nu_0}{
\nu'_0}\equiv a(f_{\rm V} \Te)
\end{align}
with $f_{\rm V}=\left(A_{21}\,\nu'_0/[A'_{21} \nu_0]\right)^2=(\aEM'/\aEM)^{-6}$, where we used the fundamental constant scaling $A_{21}/\nu_0\propto \aEM^3$. This relation allows one to trade $a$ in terms of $\Te$.
Preparing the tables over
$\tauS$, $\etac$ and $\Te$, we can then obtain scaled versions as
\begin{align}
P^{1\rm D}_{\rm esc}(\tauS, \etac, \Te)
\rightarrow
P^{1\rm D}_{\rm esc}(\tauS', f_{\etac} \etac', f_{\rm V} \Te)
\end{align}
where $f_\eta=1/\sqrt{f_{\rm V}}$ restores the temperature dependence of $\etact$ and $\tauS'$ and $\etac'$ are evaluated using the scaled atomic variables. 
For the tables we use $\Te=300-10^5$~K and $\etac=10^{-8}-10^8$. For the singlet $\HeIlevel{2}{1}{P}{1}-\HeIlevel{1}{1}{S}{0}$ line, we use $\tauS=10^{-5}-10^{10}$, while for the triplet $\HeIlevel{2}{3}{P}{1}-\HeIlevel{1}{1}{S}{0}$ resonance $\tauS=10^{-5}-10^{3}$ suffices.\footnote{This is motivated by values that are encountered for the standard recombination problem but then scaled for a range of $\simeq 20\%$ $\alpha$-variations.}
For $\tauS\leq 10^{-5}$, linear extrapolation is applied. The grid is setup in log-space with a density of 10 to 15 points per decade. A four-point Lagrange polynomial interpolation is performed in each dimension.
Indeed we find that this procedure works extremely well (usually to better than 0.1\% precision) over a wide range of the parameters. 

\begin{figure}
    \centering
    \includegraphics[width=\linewidth]{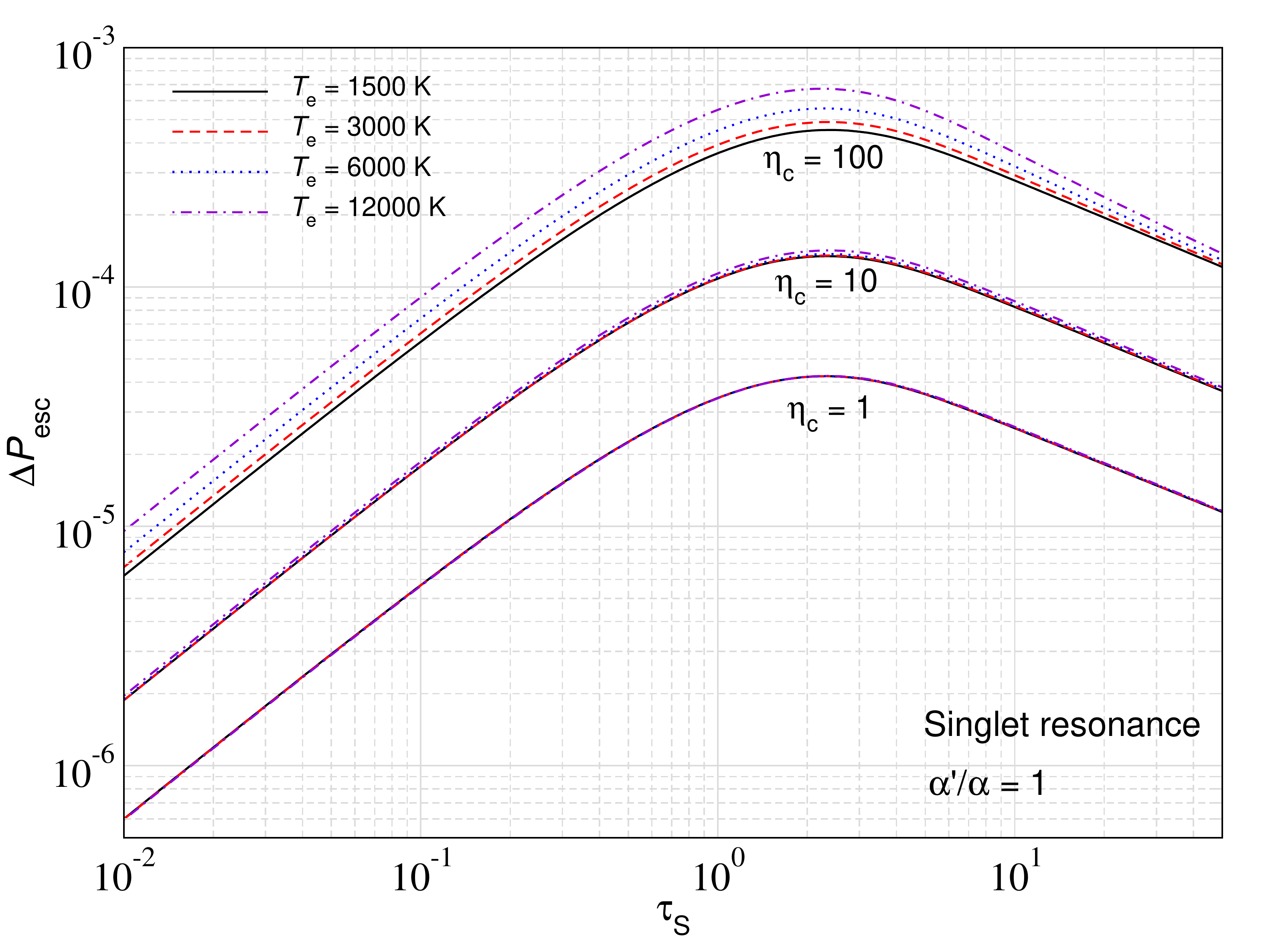}
    \\
    \includegraphics[width=\linewidth]{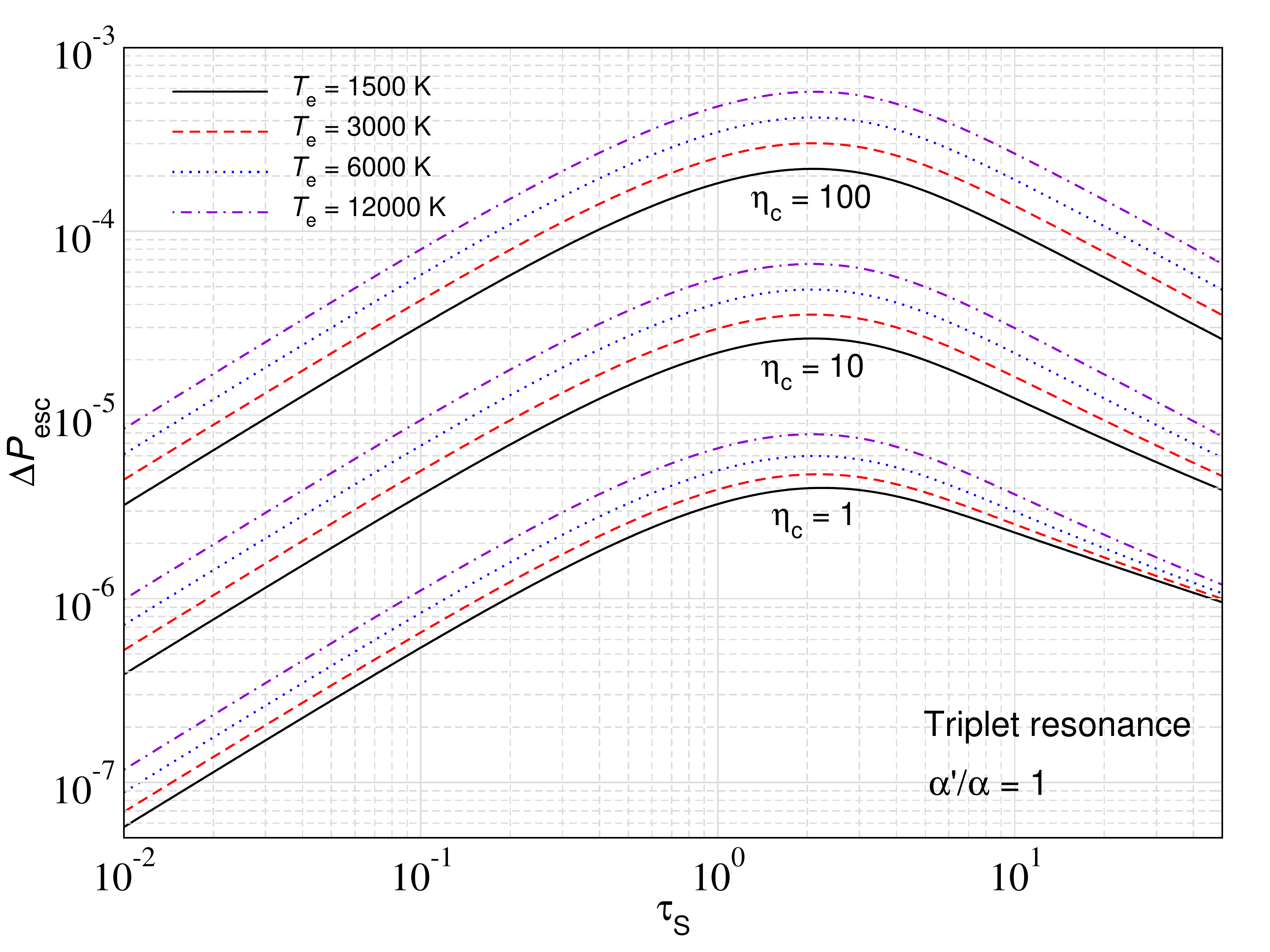}
    \\
    \caption{Dependence of $P^{1\rm D}_{\rm esc}$ for the singlet and triplet lines on $\Te$, $\etac$ and $\tauS$ for the standard value of $\aEM$ (see text for discussion). The results were obtained by direct integration, but are accurately represented by our tabulation scheme.}
    \label{fig:DPesc_LCDM}
    \vspace{-3mm}
\end{figure}
\begin{figure}
    \centering
    \includegraphics[width=\linewidth]{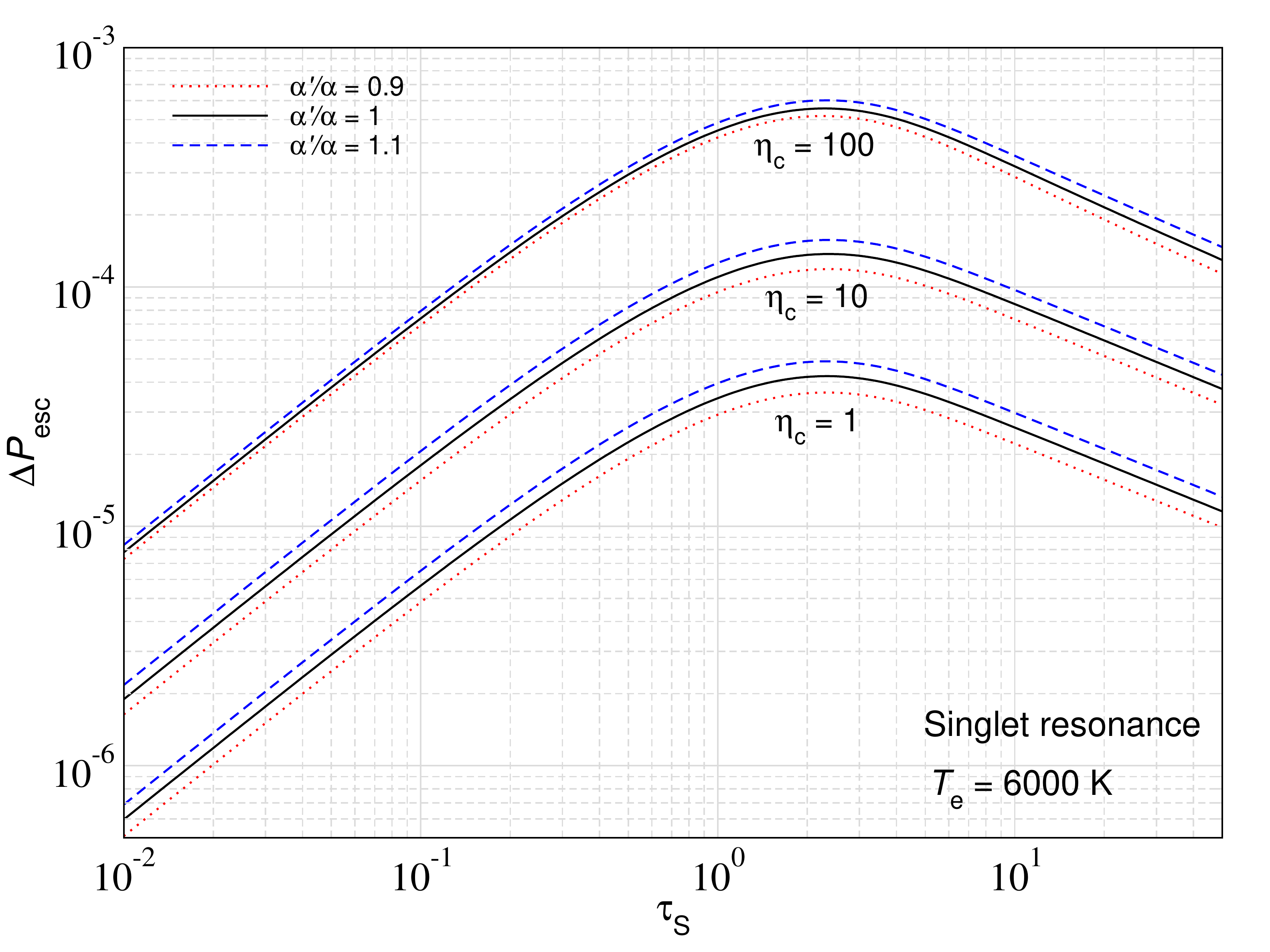}
    \\
    \includegraphics[width=\linewidth]{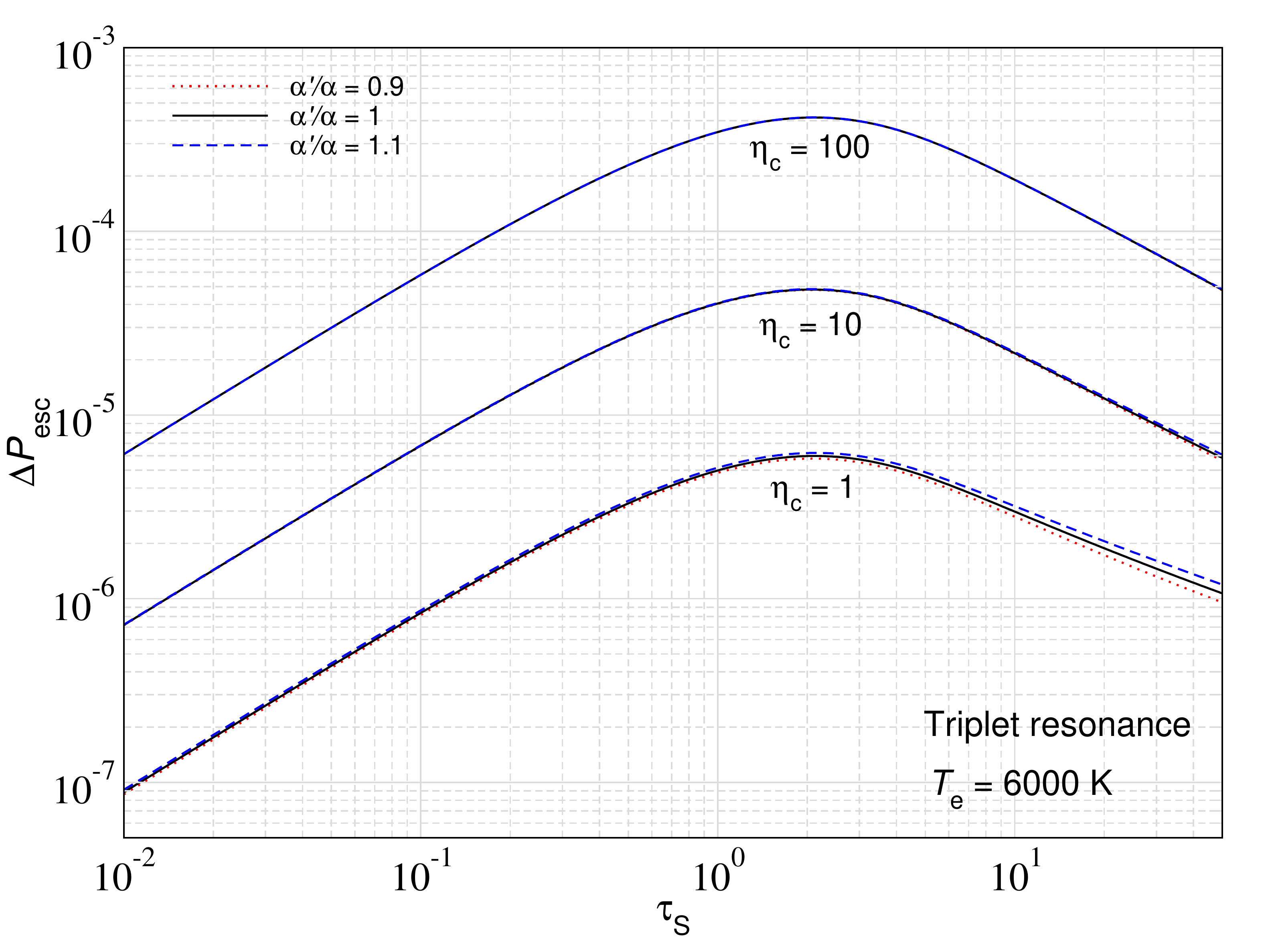}
    \\
    \caption{Dependence of $P^{1\rm D}_{\rm esc}$ for the singlet and triplet lines on $\aEM'/\aEM$, $\etac$ and $\tauS$ for the standard value of $\aEM$. The values of $\tauS$ and $\etac$ are directly evaluated for the scaled atomic species and therefore do not directly modify the escape probability. In this sense, all the visible changes are due to changes of the Voigt-parameter $a'/a\propto (\aEM'/\aEM)^{-3}$. The results were obtained by direct integration, but are accurately represented by our tabulation scheme.}
    \label{fig:DPesc_var_alpha}
\end{figure}

In Fig.~\ref{fig:DPesc_LCDM}, we illustrate the dependence of $P^{1\rm D}_{\rm esc}$ for the singlet and triplet lines on $\Te$, $\etac$ and $\tauS$ for the standard value of $\aEM$. The temperature dependence decreases with $\etac$, as expected from the fact for $\etac\rightarrow 0$ the dependence on $a=a(\Te)$ and $\Te$ drops out of the expression in Eq.~\eqref{eq:DPesc_chi_1D}. For the triplet line, even when $\etac\simeq 1$ a significant dependence on $\Te$ can be observed, while for the singlet case the corresponding curves become practically independent of $\Te$. As we will see below (Fig.~\ref{fig:DPesc_var_alpha}), the escape probability of the triplet line is practically independent of $a$, such that the main temperature dependence only enters through $\etact=\etac\,\Delta\nu_{\rm D}/\nu_0$, while for the singlet line also $a$ matters.

At $\tauS\ll1$, a quasi-linear scaling with $\tauS$ is found (see Fig.~\ref{fig:DPesc_LCDM}), which directly follows when performing a Taylor-series expansion of  Eq.~\eqref{eq:DPesc_chi_1D} to linear order in $\tauS$:
\begin{align}
  \label{eq:DPesc_chi_1D_corr}
  \Delta P^{1\rm D}_{\rm esc}
  &\approx \tauS \int_0^1 \!\!\!\id\chi\!
  \left\{1-\chi-
  \frac{1-\expf{-\tauct(\chi)[1-\chi]}}{\tauct(\chi)}\right\}
\end{align}
and justifies the adopted extrapolation procedure for $\tauS\rightarrow 0$.

To illustrate the dependence of the escape probability on the fundamental constants, we note that the escape integral has no explicit dependence on $\me$, but only indirectly from the changes of $\tauS$ and $\etac$ at fixed physical parameters like temperature and density. The only explicit dependence enters due to modifications of $a$. This is shown in Fig.~\eqref{fig:DPesc_var_alpha}. The singlet line exhibits noticeable changes for all shown values of $\etac$, while the triplet line is practically independent of $a$.
This arises from the much smaller typical value of $a$ for the triplet line and associated dominance of the Doppler core in the escape problem \citep[see Fig.~B.1 of][for some illustrations of the integrand]{Jose2008}, which makes the escape problem for the triplet line less dependent on $a$.
For the singlet line we find $\Delta P^{1\rm D}_{\rm esc}\propto (\aEM'/\aEM)^{1.5}$ for the shown cases.
Our tabulation scheme described above captures all dependencies accurately.

\subsection{Effects on the recombination lines}\label{sec:linesVFC}
\begin{figure}
    \centering
    \includegraphics[width=\linewidth]{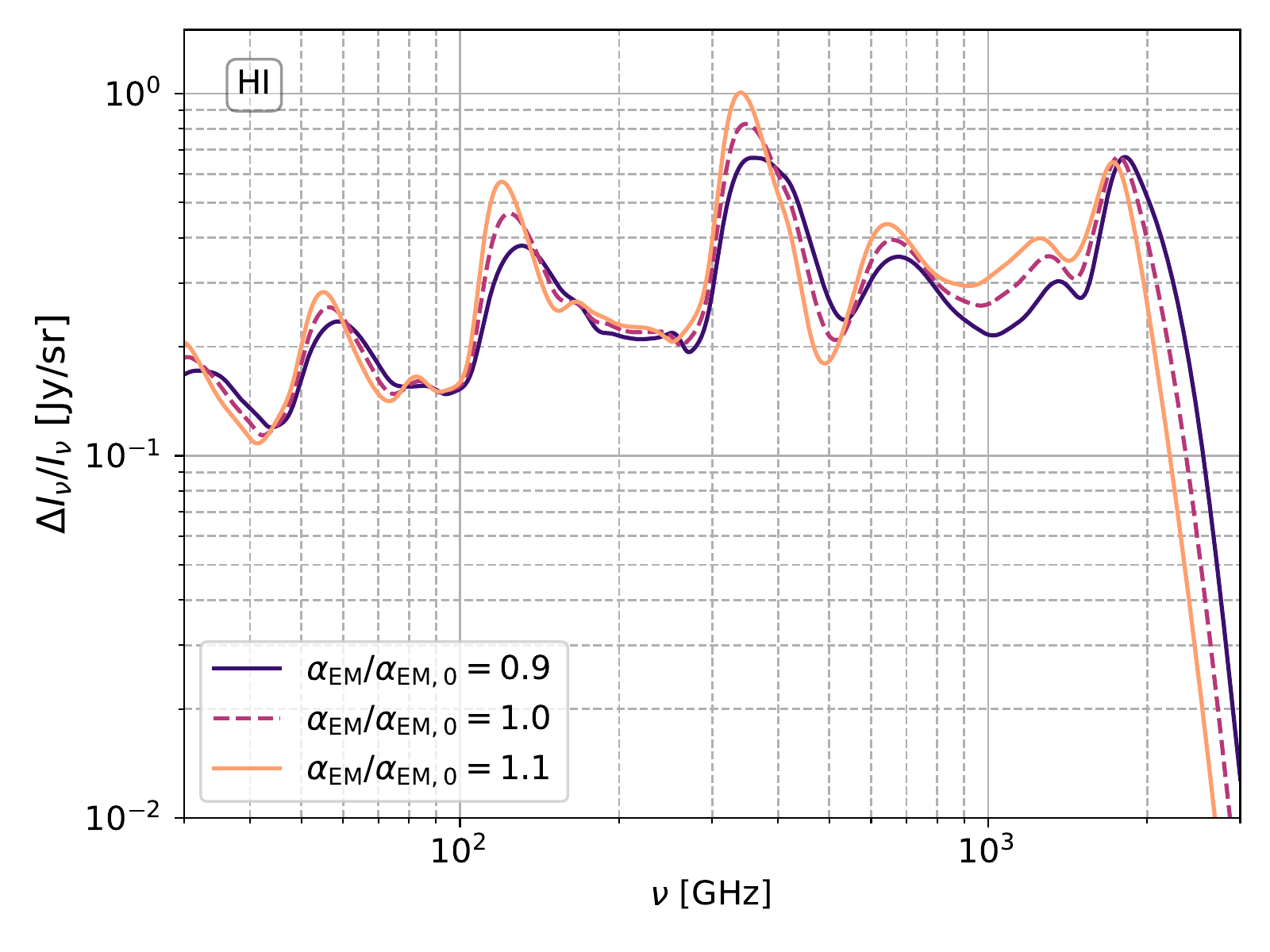}
    \\[-0.5mm]
    \includegraphics[width=\linewidth]{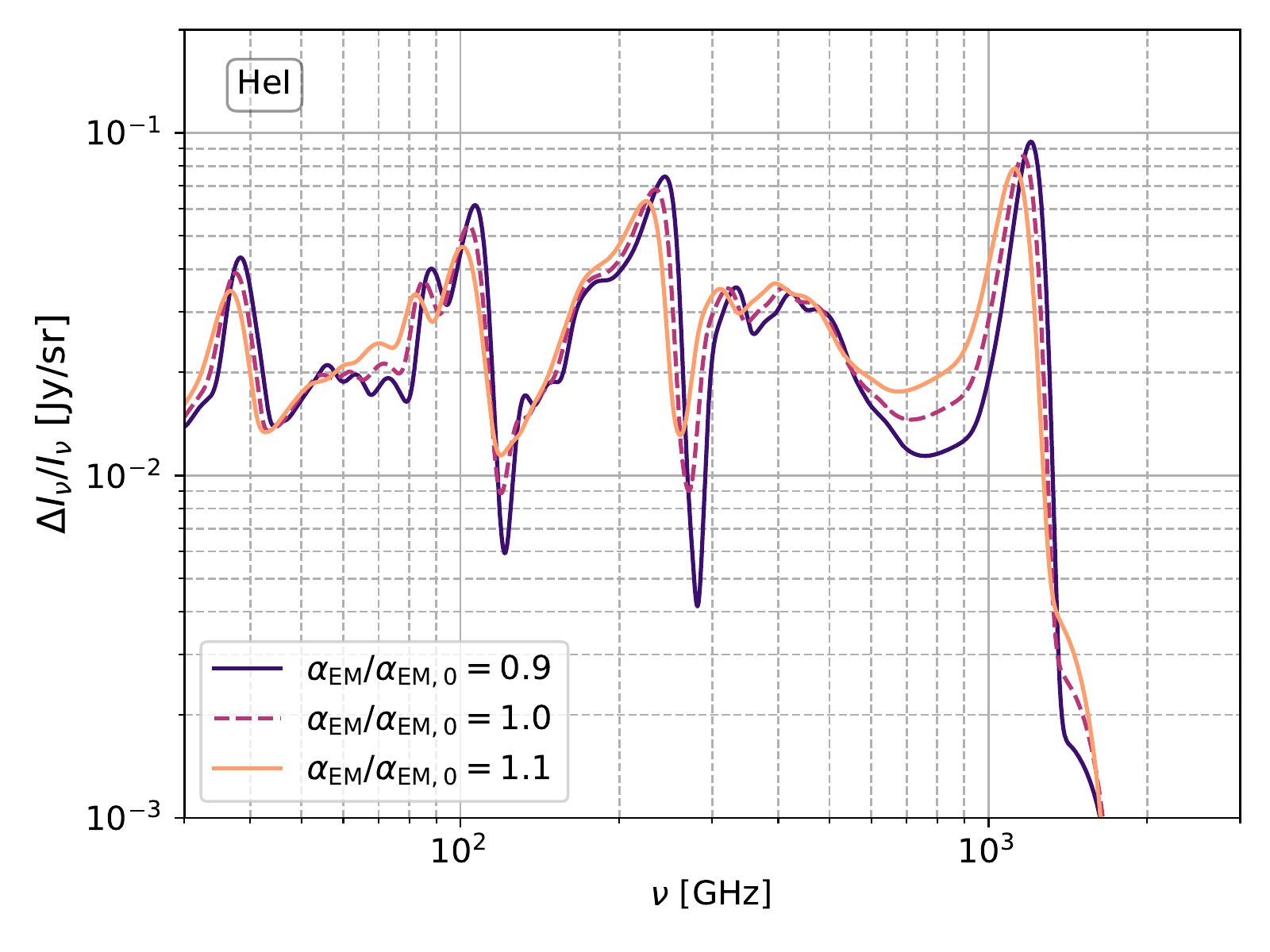}
    \\[-0.5mm]
    \includegraphics[width=\linewidth]{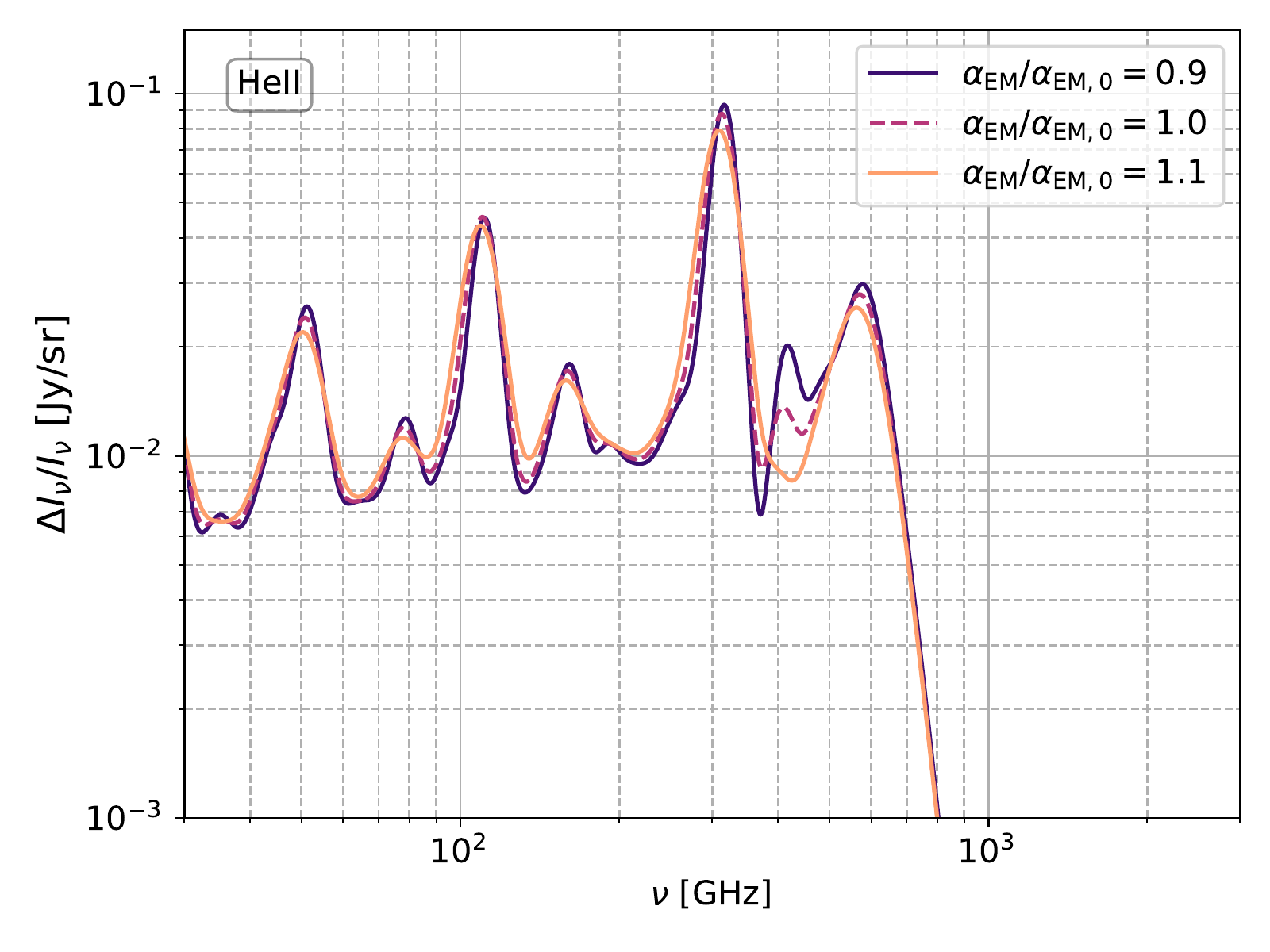}
    \\[-0.5mm]
    \caption{Changes in the cosmological recombination lines due to variations in $\aEM$ for the values $\aEM/\aEMs = \{0.9,1.0,1.1\}$. The dashed curve is shown for the \LCDM case ($\aEM/\aEMs = 1.0$). This has been shown for the hydrogen lines \emph{(top)}, singly ionised helium \emph{(middle)} and doubly ionised helium \emph{(bottom)}.}
    \label{fig:vfcCrr}
\end{figure}

The impact of varying the fine structure constant on the cosmological recombination lines is shown in Fig.~\ref{fig:vfcCrr}. As in Sec.~\ref{sec:DPesc}, the fiducial value is compared to the two cases with $\Delta\aEM/\aEM = \pm \,0.1$. In Fig.~\ref{fig:vfcCrr}, the modified recombination lines are shown for HI \emph{(top)}, HeI \emph{(middle)} and HeII \emph{(bottom)}. 
For hydrogen, the net impact of a larger $\aEM$ is an amplification on the spectral features. We find that the high frequency peaks roughly change as $\propto\aEM^2$. Given the primary recombination mechanism described in Sec.~\ref{sec:recap_vfc}, with an acceleration of recombination expected for increasing $\aEM$, this makes sense; however, the detailed effects are quite subtle, where the impact on recombination is a complicated combination of line enhancements versus broadening. The latter leads to a redistribution of photons across the CRR and hence reduction of emission in some bands, with the width being directly linked to the relative duration of the recombination process. We also note that at low frequencies (not shown here), we do not see any significant change in the amplitude of the distortion, aside from some modifications from the free-free absorption process. Given that at low frequencies no spectral features are visible, this shows that a net redistribution of the emission in different bands occurs while leaving the total number of photons added nearly constant. Thus visible changes to the CRR only appear where individual lines can be identified. 

For the helium lines, additional complications arise from the changes to the importance of electrons scattering. With increasing $\aEM$ the effective scattering $y$-parameter increases and so does the smearing of the lines. This is indeed visible in the HeII spectrum. We can notice that the HeII lines remain almost constant in amplitude but mainly change their width. This highlights that for the total spectrum it is hard to understand how the effects propagate into the final CRR just from the simple scaling of variables given above.

An increase in $\aEM$ also causes a small net drift of the main lines for all species of distortion to lower frequencies. However, the effect is much smaller than what would be guessed from the energy scaling of the transition frequencies, $\nu_{ij}\propto \aEM^2$. In reality, the recombination process also occurs at roughly $\aEM^2$ times higher redshifts, leaving the ratio $\nu_{ij}/(1+z_{\rm em})$ roughly constant for individual transitions. Nevertheless, the positions of the lines are tracers of when the recombination process happened. Interestingly, the shifts are more noticeable for the HI and HeI contributions than for HeII, for which line broadening effects seem to be more pronounced.

For larger $\aEM$ we can also see added spectral peaks emerging. Examples are at $\nu\simeq80{\rm GHz}$ and $\nu\simeq160{\rm GHz}$ in the HI contribution. This has to do with the way the emission from various transitions overlaps and compensate each other, partially canceling or interfering constructively.
For HeII (lower part of Fig.~\ref{fig:vfcCrr}), larger values of $\aEM$ more strongly smear the overall shape, with peaks being compressed and troughs being raised. In this case, we also notice a deterioration of the peak feature at $\nu\simeq450{\rm GHz}$, indicating modifications in the relative importance of various transitions, here related to the HeII Balmer lines.

\begin{figure}
    \centering
    \includegraphics[width=\linewidth]{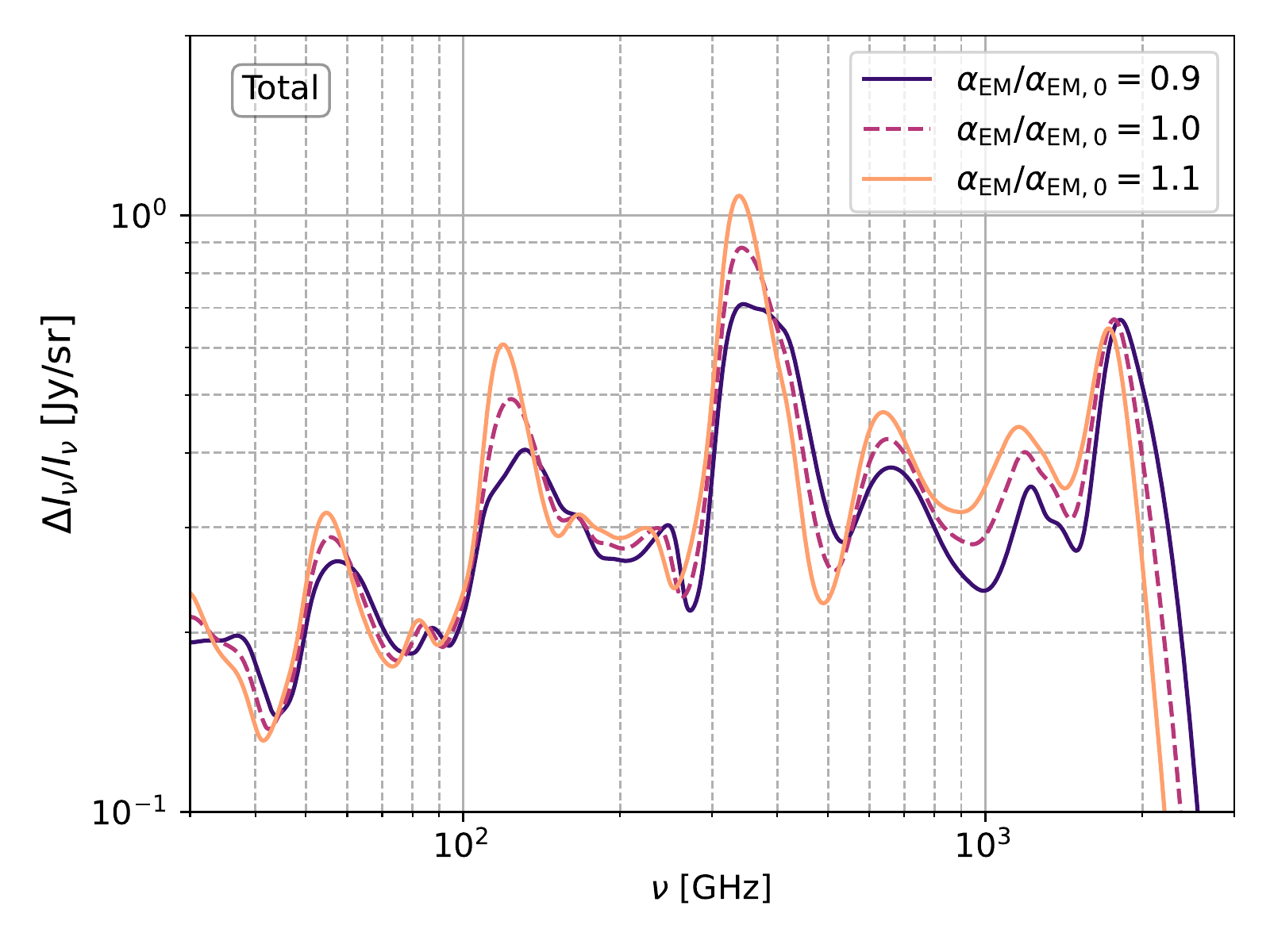}
    \includegraphics[width=\linewidth]{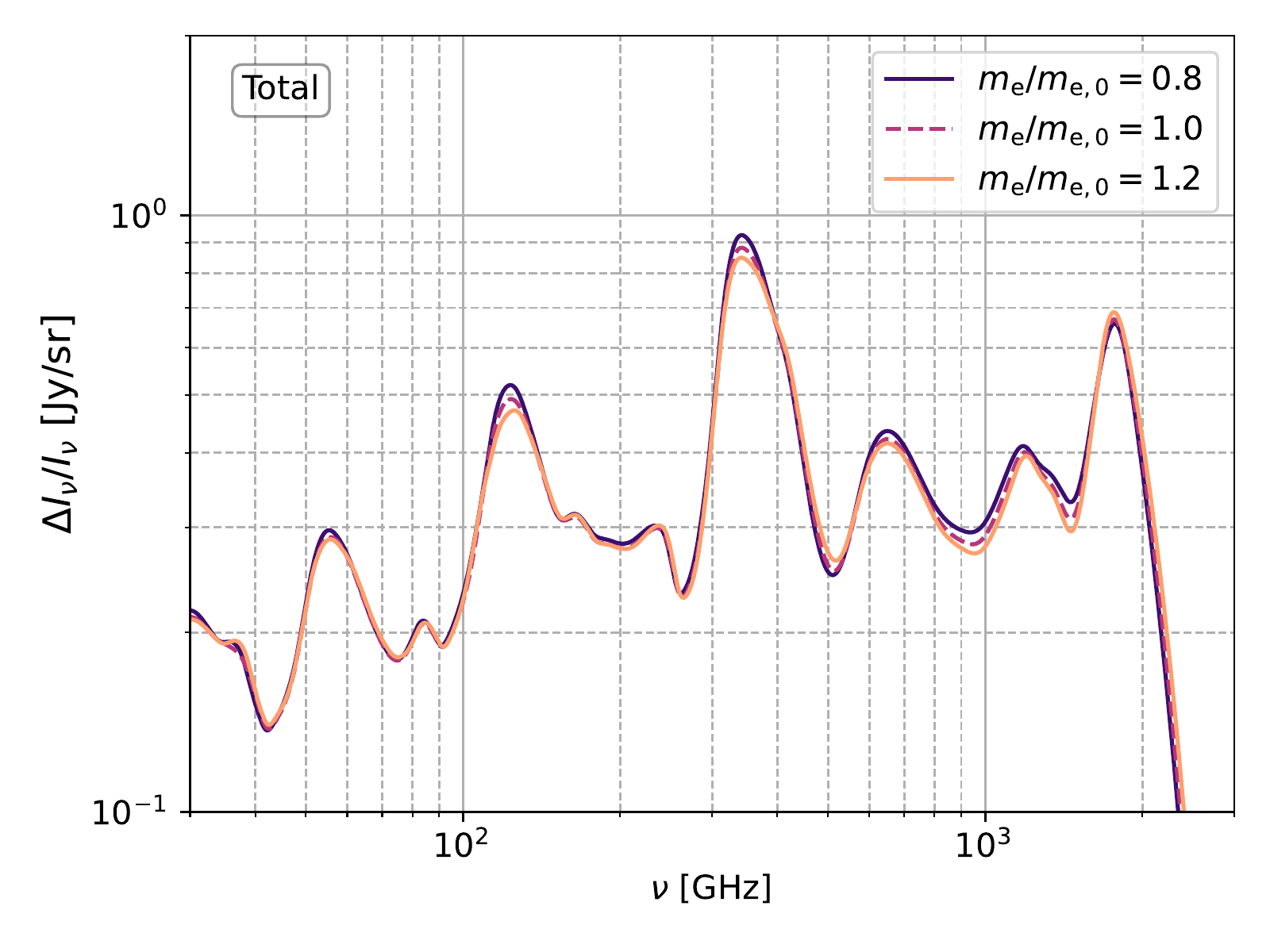}
    \caption{The total impact on the recombination lines from the variations shown in Fig.~\ref{fig:vfcCrr} for $\Delta\aEM/\aEMs\pm0.1$ \emph{(top)}. The total variations due to changes in $\Delta\me/\mes\pm0.2$ are shown for comparison \emph{(bottom)}.}
    \label{fig:vfcCrrTotal}
\end{figure}

When all three atomic species are combined into a single distortion, the result is as shown in Fig.~\ref{fig:vfcCrrTotal}. Here the more intricate variations due to helium are smeared out by the larger magnitude signal from hydrogen, as was the case for early dark energy. However, this still provides richer structure on the inter-level emission bumps throughout the profile. This should generally allow one to distinguish $\aEM$ variations in the CRR from other parameters using high resolution spectrometers.

For comparison, in Fig.~\ref{fig:vfcCrrTotal} we also show the same spectra but for changes to $\me$. Although we have doubled the variation to $\Delta\me/\mes=\pm0.2$, one can see that the changes of the total CRR are a lot smaller. By decreasing the electron mass one can observe a small overall enhancements of the spectral features.
Therefore, $\me$ will be harder to detect with a CMB spectrometer, with the largest response being visible at about $\simeq 1$~THz. 
We confirmed numerically that this has to do with the precise interplay of the emissivities and energies scaling, leaving the main effect to be a small change in the width of the spectral features.
This directly contrasts with the influence that $\me$ has on the CMB anisotropies as previously discussed in the literature \citep{Hart2017,Hart2020a}. However, here the observable is related to the Thomson visibility function, which has a different non-canceling dependence on $\me$.

\subsubsection{Relationship between $\aEM$ and $\TCMB$}\label{sec:alphaT}
\begin{figure}
    \centering
    \includegraphics[width=\linewidth]{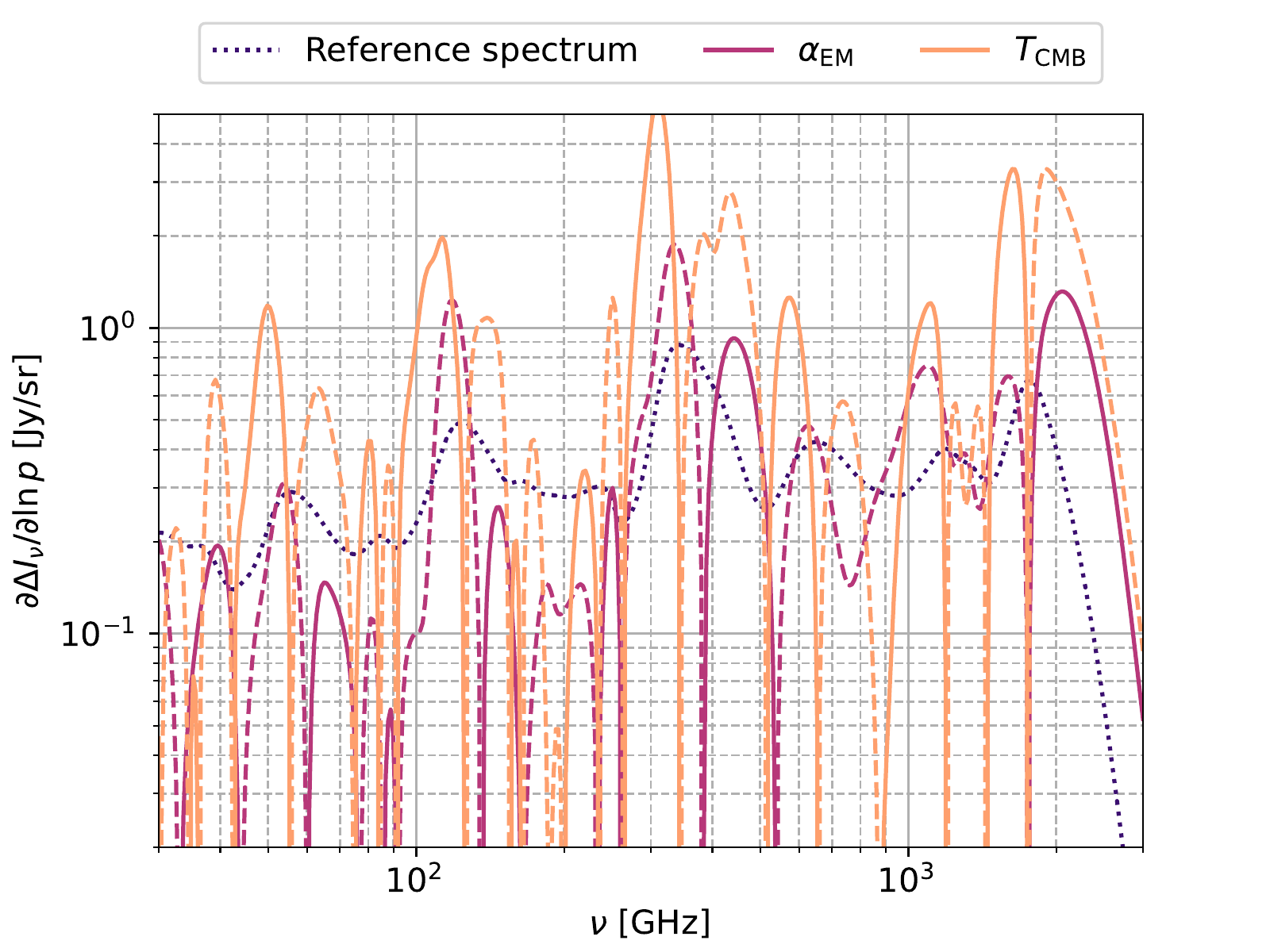}
    \\
    \caption{Weighted derivatives for changes in the recombination lines for variations of $\aEM$ vs. variations in $T_{\rm CMB}$. The negative derivatives are indicated by dashed lines. Both are compared against the $\Lambda$CDM reference spectra  to see how the derivatives compare against the full spectral signal. Similarly, we have used the $\ln p$ weighting for the denominator as this will more appropriately compare to the fiducial CRR.}
    \label{fig:dlnI_alpha_T}
\end{figure}

The fine structure constant has very clear degeneracies with the monopole temperature of the CMB, $T_{\rm CMB}$ due to the similar effects they have on the last scattering surface position. In Fig.~\ref{fig:dlnI_alpha_T}, we show the relative derivatives for $\aEM$ \emph{(purple)} and $\TCMB$ \emph{(orange)}. The reference CRR spectrum for hydrogen and helium combined is shown as a dotted line. The large-scale features of the derivatives spectrum align in most cases: for example the positive-negative envelope in $\aEM$ at $\nu\simeq1800$ GHz, close to the edge of the Lyman-continuum, is mirrored at a higher amplitude in $\TCMB$. Similarly the mirror in the derivatives is also present between $\aEM$ and $\TCMB$ in the predominantly bound-bound transitions\footnote{Though the free-bound radiation is present, it is void of the features that we typically associate with bound-bound transitions at $\nu\lsim100$ GHz.} at $\nu\lsim 100$ GHz. 
The remainder of the spectra, in the mid-range of the proposed bands by \vtwenty, show distinct differences between the responses from the two parameters. It was known from previous studies that $\aEM$ and $\TCMB$ had some degeneracy breaking effects due to $\aEM$ affecting the recombination physics while not affecting the global expansion rate (under the current assumptions). In comparison, $\TCMB$ affects the positioning of the \emph{matter-radiation equality} epoch as well. This leads to a wider impact at higher redshifts \cite[see][for more details]{Hart2020a}. With the full deviations implicitly found in the interaction rates and the conductance variations (see Sec.~\ref{sec:recap_vfc}), the changes that will primarily affect the Lyman, Balmer and higher order series will be present in the $\aEM$ derivatives and less obvious in the $\TCMB$ responses. As originally shown in \citet{Chluba2008T0}, the generic effect from changing $\TCMB$ is to shift the peaks of the CRR back and forth. Therefore, the interaction rates have a more complex effect from variations in $\aEM$.

\subsubsection{Relationship between $\me$ and $\Neff$}\label{sec:meNeff}
\begin{figure}
    \centering
    \includegraphics[width=\linewidth]{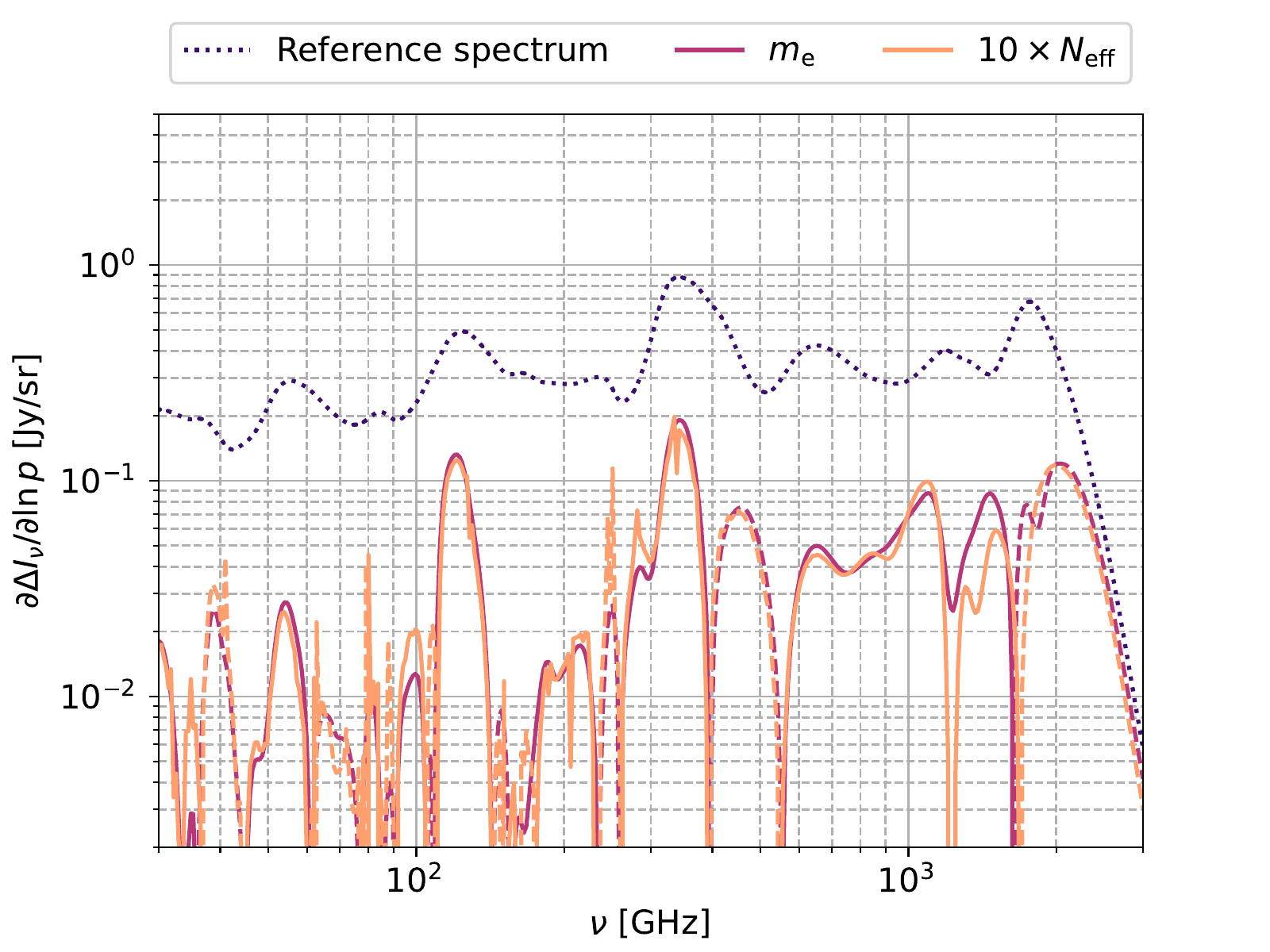}
    \\
    \caption{Recombination radiation derivatives for the electron mass rescaling against the $\Neff$ derivatives originally shown in \citet{Hart2020c} (multiplied by a factor of 10 for readability). These have been weighted by the variations in parameter $\delta p/p$. }
    \label{fig:dlnI_me_Neff}
\end{figure}

The covariance that can be highlighted otherwise is the relationship between $\me$ and $\Neff$, the relativistic degrees of freedom. The weighted derivatives are shown in Fig.~\ref{fig:dlnI_me_Neff}. Specifically the negative and positive spectra for $\me$ and $\Neff$ almost perfectly emulate each other; however, there are small structural changes in the inter-spectral gaps. This is due to the more complicated atomic variations from $\me$ compared with the broader effects on the expansion rate associated with $\Neff$. When looking at the finer structure of the spectral responses, differences can be ascertained as the resolution and sensitivity increase. For example, at $\nu\simeq 1500$ GHz in Fig.~\ref{fig:dlnI_me_Neff}, the broader changes in the helium feedback lines for $\me$ are different to $\Neff$. Therefore, we would expect that a sufficiently high sensitivity would distinguish electron mass from changes to the relativistic degrees of freedom. 

\changeJ{We also note that the changes from $\Neff$ are numerically harder to compute, given that very large changes are required to see any responses \citep{Hart2020c}. This is the origin of some numerical imperfections, visible as kinks in Fig.~\ref{fig:dlnI_me_Neff}, however, they do not affect the main conclusions, in particular once external priors are added.}

\subsection{Possible detections for $\aEM$ and $\me$ with the CRR}\label{sec:vfcFisher}
In this section, we use the spectral variations in the CRR caused by a change in $\aEM$ and $\me$ to test the constraining strength for future spectrometer missions. This involves a Fisher forecast where one additional parameter ($\aEM$, $\me$) modifies the standard \LCDM matrix. This will be shown for the following variety of assumptions for the priors on the standard parameters: no priors and priors from \planck 2018.
The contours in this section were generated using {\tt GetDist} as in Sec.~\ref{sec:forecastEDE} using Gaussian random samples. However in this section, we use the previously generated \planck chains to add inverse covariances to the Fisher matrix \citep{Hart2020a}.

\begin{figure}
    \centering
    \includegraphics[width=\linewidth]{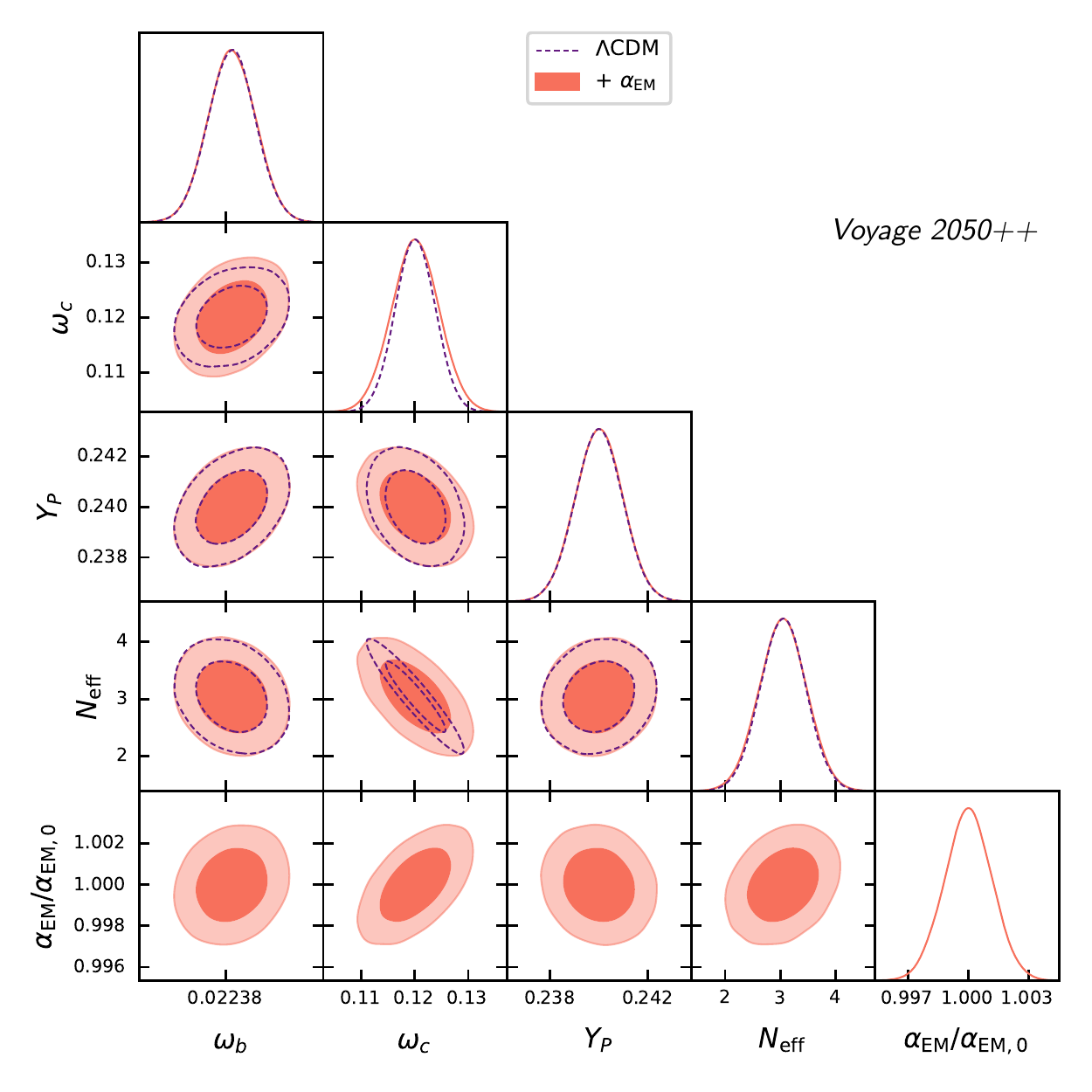}
    \\
    \caption{Rudimentary contour plot for the explored parameters from \citet{Hart2020c} with $\aEM$ added into the analysis. The contours are compared for the futuristic configuration \vtwentypp with and without the added fundamental constants.}
    \label{fig:alphaFisherCRR}
\end{figure}
\begin{figure}
    \centering
    \includegraphics[width=\linewidth]{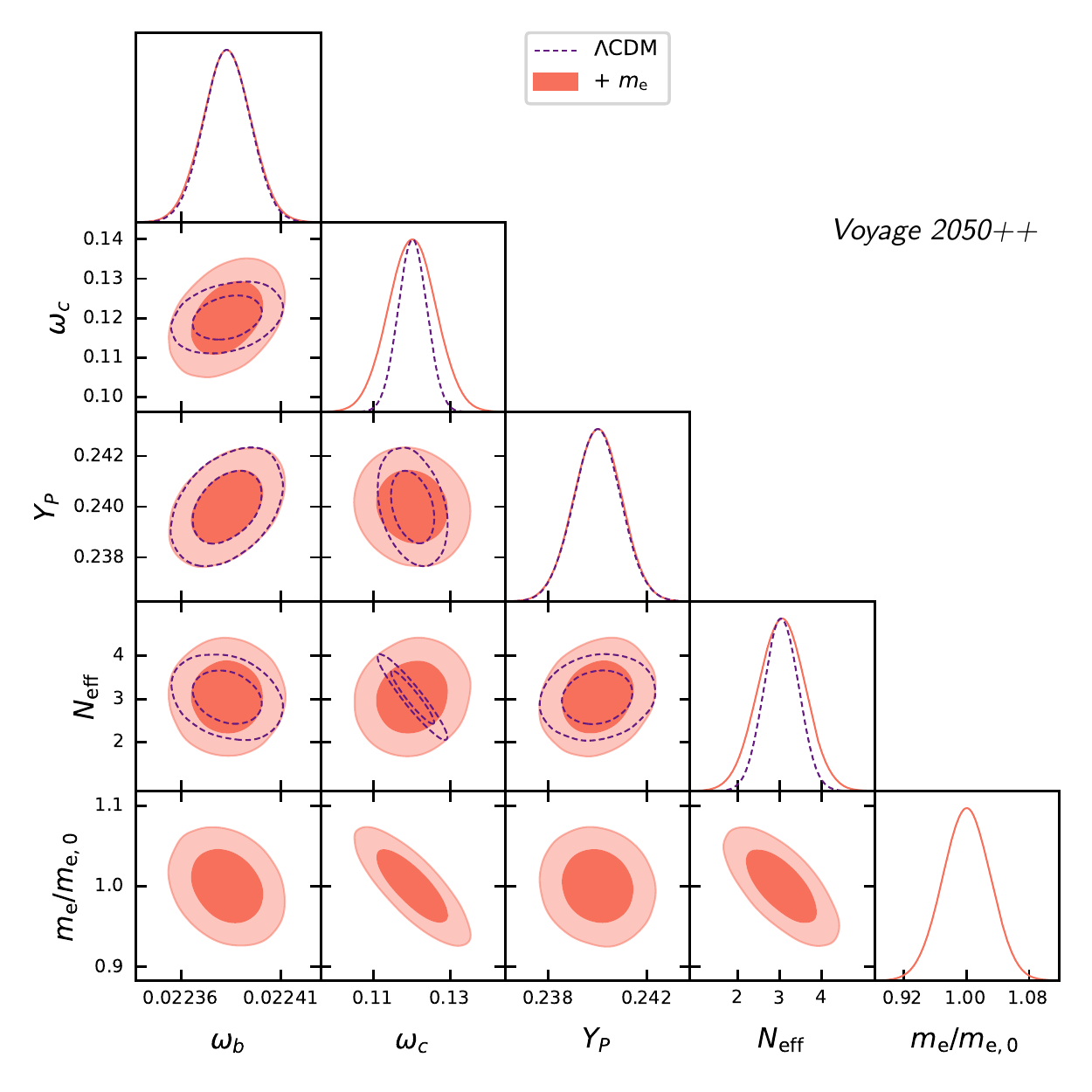}
    \\
    \caption{The same forecasted posterior contours as in Fig.~\ref{fig:alphaFisherCRR} however for variations in electron rest mass $\me$. Note that the configurations and varying parameter sets are the same as well.}
    \label{fig:meFisherCRR}
\end{figure}

\subsubsection{Case 1: Spectrometer only}\label{sec:specOnly}
The most simple case for this is where we forecast the detectability with a futuristic spectrometer such as \vtwentypp. For this we use the same noise profiles as described in Sec.~\ref{sec:forecastEDE} for \vtwenty. Specifically, we will use the standard Voyage setup (\vtwenty) and $50\times$ higher \changeJ{sensitivity} (\vtwentypp) to highlight the potential of future CMB spectrometers as a cosmological probe. 

The comparison for \vtwentypp for \LCDM and adding $\aEM$ is shown in the contours of Fig.~\ref{fig:alphaFisherCRR}.
When varying $\aEM$ the contours show a general broadening once the fine structure constant is added to the analysis. This softens the direct degeneracy between $\omc$ and $\Neff$ that we have previously discussed in this paper and previous works \citep{Hart2020c}, transferring it to correlations with $\aEM$. However, the contours are not altered to a great amount in general; this is owed to the largely distinctive and unique spectral patterns associated with $\aEM$ that we have discussed in Sec.~\ref{sec:linesVFC}. The standard deviations for this configuration with $\aEM$ are shown in  Table~\ref{tab:alphaCRR}. From the data in this table, and the previous findings from \citet{Hart2020c}, we can see that $\omc$ and $\Neff$ are far too weakly constrained by \vtwenty alone, requiring higher sensitivity and complimentary probes to reach competitive results.\footnote{We ask the reader to bear in mind that the errors for $\Neff$ in Table~\ref{tab:alphaCRR} and Table~\ref{tab:meCRR} are indeed large compared to modern probes; however, they are quoted here to underpin the potential of CRR oriented spectrometers when the sensitivity can be reached.} As $\aEM$ is included in the analysis, we can see from Table~\ref{tab:alphaCRR} that the errors on the parameters do not expand, save for a $\simeq 19\%$ increase for $\omc$.

\begin{table}
    \centering
    \begin{tabular}{c|c|c|c}
    \hline\hline
        Parameter & \vtwenty & \vtwenty & \vtwentypp \\
         & \LCDM & + $\aEM$ & + $\aEM$ \\
        \hline
        $\omb$ & $0.00057$ & $0.00058$ & $0.00002$ \\
        $\omc$ & $0.186$ & $0.222$ & $0.004$ \\
        $\Yp$ & $0.0478$ & $0.0481$ & $0.0010$ \\
        $\Neff$ & $20.40$ & $21.30$ & $0.43$ \\
        $\aEM$ & \textemdash & $0.059$ & $0.001$ \\
        \hline\hline
    \end{tabular}
    \caption{Initial covariance forecasts from \vtwenty and \vtwentypp spectrometers when the parameters listed here are modified, for varying $\aEM$. In this case, the \vtwentypp configuration is $50\times$ more sensitive than the standard \vtwenty.}
    \label{tab:alphaCRR}
\end{table}

\begin{table}
    \centering
    \begin{tabular}{c|c|c|c}
    \hline\hline
        Parameter & \vtwenty & \vtwenty & \vtwentypp \\
         & \LCDM & + $\me$ & + $\me$ \\
        \hline
        $\omb$ & $0.00057$ & $0.00059$ & $0.00001$ \\
        $\omc$ & $0.186$ & $0.309$ & $0.006$ \\
        $\Yp$ & $0.0478$ & $0.0481$ & $0.0010$ \\
        $\Neff$ & $20.40$ & $27.90$ & $0.56$ \\
        $\me$ & \textemdash & $1.53$ & $0.030$ \\
        \hline\hline
    \end{tabular}
    \caption{Electron mass $\me$ forecasted covariances, $\sigma_{ij}$, from \vtwenty and \vtwentypp spectrometers when these parameters are modified. The configurations are the same as the ones in Table~\ref{tab:alphaCRR}.}
    \label{tab:meCRR}
\end{table}

For $\me$, the situation is slightly more complicated as the contours in Fig.~\ref{fig:meFisherCRR} show. Both $\omc$ and $\Neff$ distributions are broadened and the effect from $\me$ all but removes the $\omc-\Neff$ degeneracy. In this case, the co-varying power is passed to the $\me$-$\Neff$ relation, as the $\me$ variations partially emulate an acceleration in the radiation era. This is corroborated by the results in Table~\ref{tab:meCRR} where the error on $\omc$ increased by $66\%$ and $\Neff$ by $36\%$ for \vtwenty.

We note that none of the other \LCDM parameters matter for the CRR and that the value of $T_{\rm CMB}$ will be measured to very high precision using the blackbody part. Of course the precludes cosmologies with varying temperature-redshift relation, which would run into several other issues of course \citep{Chluba2014TRR}.

\subsubsection{Case 2: Spectrometer with Planck 2018}\label{sec:specPlanck}

\begin{figure}
    \centering
    \includegraphics[width=\linewidth]{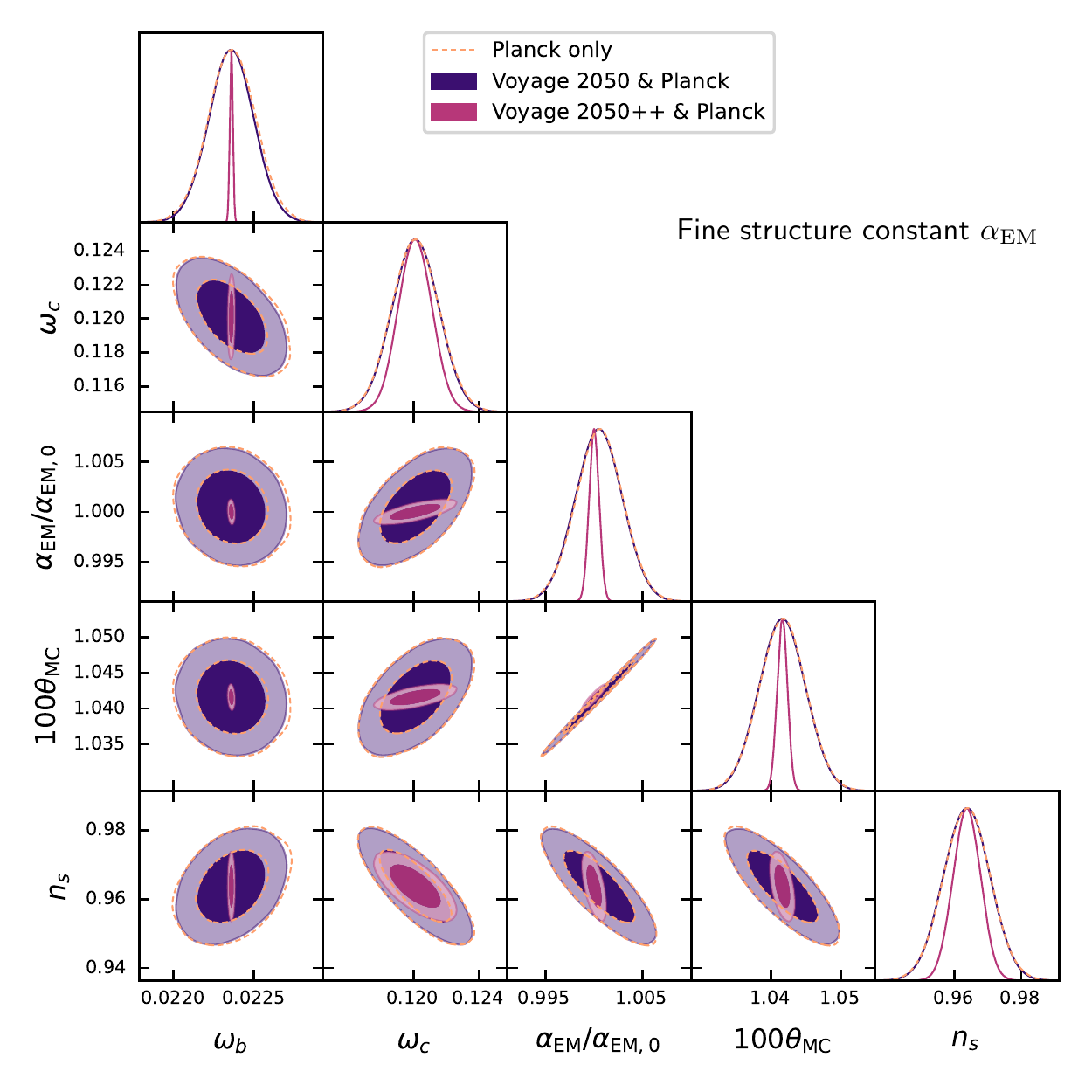}
    \\
    \caption{Forecasted contours for variations in $\aEM$ that correspond to the results in Table~\ref{tab:alphaForecast}. The emphasised parameter set $\{\omb,\omc,\aEM,\thetaMC,\ns\}$ show the most important variations when the fine structure constant is considered with spectral distortions. \changeL{Here the parameter drift in $\aEM$ arises from assuming a fiducial picture of $\aEM/\aEMs=1.0$ for the CRR case.}}
    \label{fig:alphaFullContour}
\end{figure}
From the spectrometer only results, the next stage is adding the \planck covariance into the analysis. Specifically, we can do this by adding the Fisher matrix $F_{ij}$ to the inverse covariance matrix from \planck ${\bf: \Sigma}_{ij}$. Note that in this case the parameters coming from \planck ($\theta_{\rm MC},\tau,\As,\ns$) have a zero contribution in the CRR Fisher matrix; we are simply investigating how the parameter errors change as the covariances are influenced.

Adding different mission configurations with \planck data for variations of $\aEM$ are shown in Fig.~\ref{fig:alphaFullContour}. The standard \LCDM parameters associated with the the power spectrum amplitude and reionisation era, $A_s$ and $\tau$ have been omitted from the plots as the addition of CRR does not alter their values significantly. The configuration for the highest sensitivity is \vtwentypp since this is the level of sensitivity in our setup where the errors start to markedly improve compared to \planck. In our forecasts with $\aEM$ and $\me$, we have only included the two energy densities: $\omb$ and $\omc$ as added free parameters since we have not included chains where $\Yp$ and $\Neff$ co-vary with the fundamental constants. Responses in $\ho$ are so small in the CRR they have not been included here \cite[see][for more details]{Hart2020c}.

Firstly, the \planck contours are shown as yellow dashed lines, with the addition of \vtwenty data in dark purple. The contours for \planck are almost identical whether one includes \vtwenty or omits it in the case of $\aEM$. However, there are some small shifts in the maximum likelihood positions. This is reflected in Table~\ref{tab:alphaForecast} where the values of the errors do not change for this configuration either. When we look at a higher precision \vtwentypp configuration, the errors start to diminish as shown in Fig.~\ref{fig:alphaFullContour}. Including $\aEM$ with \planck+\vtwentypp forecasts, the finer spectral shape of $\aEM$ is complimented by the much higher sensitivity and the error on $\aEM$ is significantly reduced. \changeJ{Some gains can even be seen for other parameters like $\ns$ due to a reduction of the $\aEM$-$\thetaMC$ contours.}

\begin{figure}
    \centering
    \includegraphics[width=\linewidth]{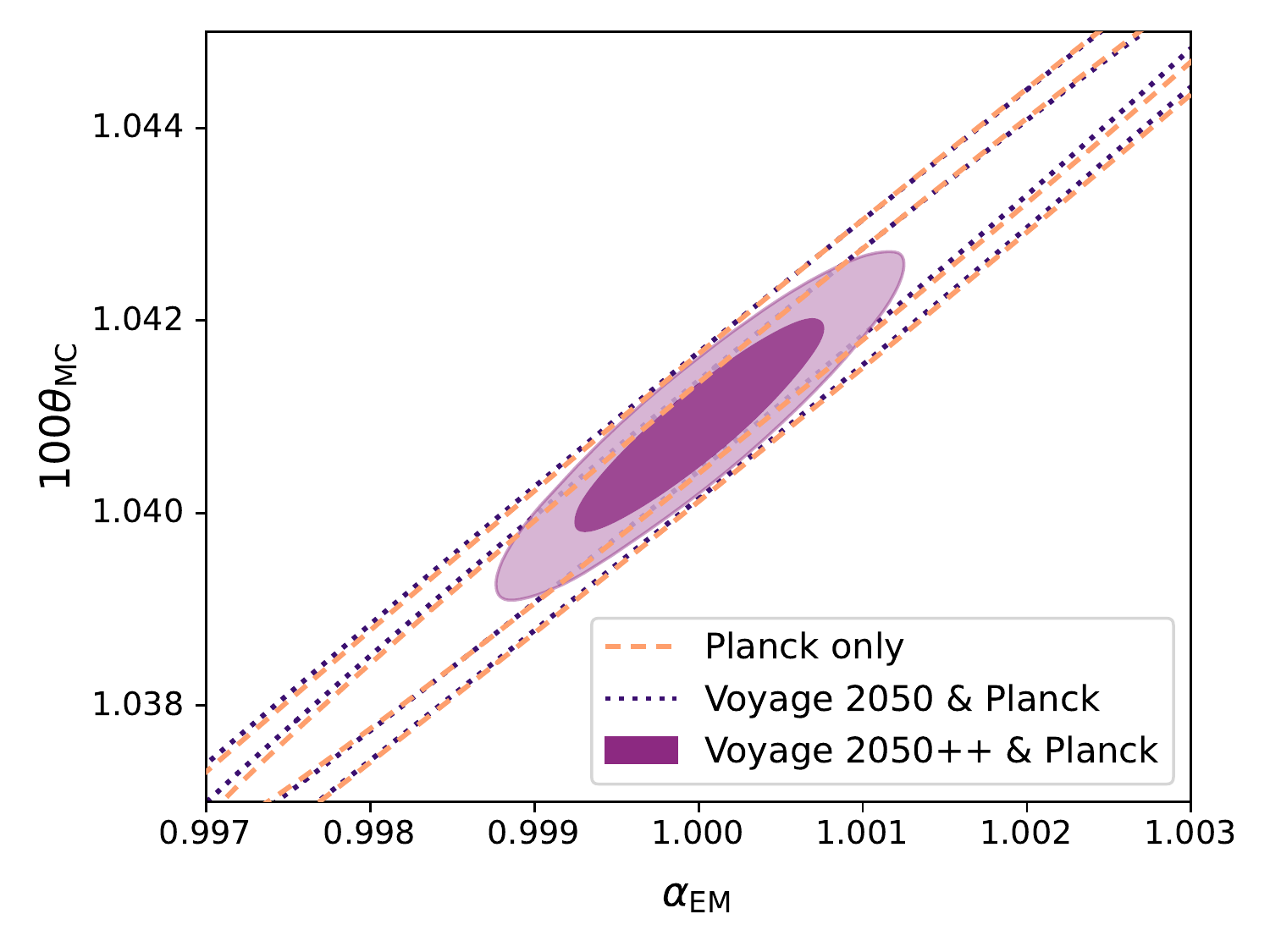}
    \\
    \caption{Single contours from Fig.~\ref{fig:alphaFullContour} specifically focused on the $\aEM$-$\thetaMC$ degeneracy to showcase the much better forecasting that \vtwentypp gives as well as the similar degeneracy line; note here that both \vtwenty and \vtwentypp have been folded in with \planck errors.}
    \label{fig:alpha2d}
\end{figure}

The degeneracy of $\aEM$ with $\thetaMC$ is not removed; however, since the parameter degeneracies with $\theta$ are not altered by other parameters, the correlations remain similar (correlation between the parameters $\xi_{\theta,\alpha} = 0.99 \rightarrow0.91$). This has been highlighted in Fig.~\ref{fig:alpha2d}. When the high precision \vtwentypp configuration is considered, even with latent degeneracies, the error on $\aEM$ is diminished to $\sigma_{\aEM} \simeq 0.0005$, which is $\simeq 5$ times smaller than \planck alone and also another factor of $\simeq 2$ better than the \vtwentypp spectrometer alone. Note that for CMB anisotropy measurements, the addition of BAO data did not change the error on $\aEM$ in previous studies \citep{Hart2020a}.

\begin{figure}
    \centering
    \includegraphics[width=\linewidth]{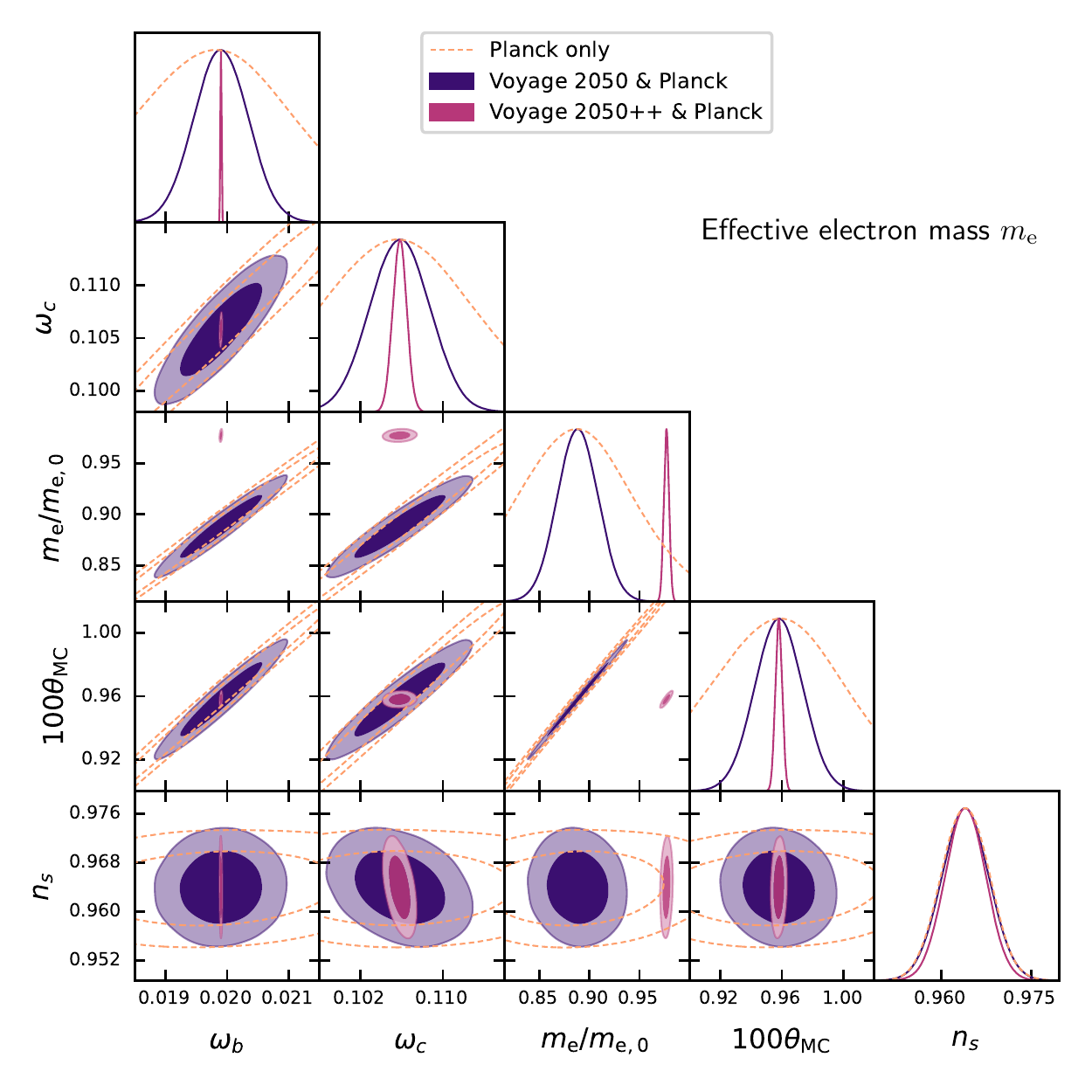}
    \caption{Forecasted contours for variations in $\me$ that correspond to the results in Table~\ref{tab:alphaForecast}. The emphasised parameter set $\{\omb,\omc,\me,\thetaMC,\ns\}$ show the most important variations when the fine structure constant is considered with spectral distortions.\changeL{As in Fig.~\ref{fig:alphaFullContour}, the shifts in $\me$ emulate the drift expected between \planck and the CRR configurations associated with the change in errors and the CRR fiducial case being $\me/\mes=1.0$.}}
    \label{fig:meFullContour}
\end{figure}
Conversely, the influence of $\me$ has a more substantial effect. In \planck 2018, the errors for $\me$ are much larger since the marginalised result is $\me/m_{\rm e, 0} = 0.888\pm0.059$ as shown in Table~\ref{tab:meForecast}.
In particular, the non-zero off-set with respect to the standard electron mass helped alleviate the Hubble tension \citep{Hart2020a}.
However, already when we combine $\vtwenty$ with \planck, we see the error in $\me$ to cascade down and move to the fiducial value: the density covariances with $\me$ are limited by the spectrometer. The error in $\me$ is already $\simeq 3$ times smaller in this case, allowing one to contest the non-zero shift in the value for $\me$ seen from \planck alone. 
\changeL{Primarily, the error on $\me$ in \planck is caused by the geometric degeneracy between $\me$ and $\thetaMC/\ho$. Since the changes in $\Delta I_\nu$ caused by variations of $\me$ are distinct from any associated with the Hubble constant, this separates the variations of the two parameters and begins to remove the correlation.} 
When we further look at \planck with \vtwentypp, the error drops by a another factor of $\simeq7$. Evidently, this very futuristic version of the \vtwenty spectrometer will give unprecedented precision on $\me$ which may help rule out a possible VFC origin of the Hubble tension.
\changeJ{We note that adding \vtwentyp to \planck (not shown here) did not lead to significant additional improvements on $\me$ since the gains on other CMB parameters did eat up some of the gains from the increased spectrometer sensitivity.}

\begin{table}
    \centering
    \begin{tabular}{c|c|c|c}
     \hline
     Parameters & \planck only  & \vtwenty & \vtwentypp \\
      & ($\mu_i \pm \sigma_i$) & \& \planck & \& \planck\\
     \hline\hline
       $\omega_b$ & $0.02236\pm0.00015$ & $0.00014$ & $0.00001$ \\
       $\omega_c$ & $0.1201\pm0.0014$ & $0.0014$ & $0.0010$ \\
       $\aEM$ & $1.0005\pm0.0024$ & $0.0024$ & $0.0005$ \\
       $100\, \theta_{\rm MC}$ & $1.0416\pm0.0034$ & $0.0034$ & $0.0007$ \\
       $\tau$ & $0.0540\pm0.0075$ & $0.0075$ & $0.0070$ \\
       $\ln(10^{10} A_s)$ & $3.043\pm0.015$ & $0.015$ & $0.014$ \\
       $n_s$ & $0.9637\pm0.0070$ & $0.0069$ & $0.0041$ \\
       \hline\hline
    \end{tabular}
    \caption{Forecasted standard deviations found for different mission configurations combining CMB anisotropies (\planck) and spectral distortions (\vtwenty, \vtwentypp). Note that here the isolated \planck parameters $\left\{\thetaMC,\tau,\logA,\ns\right\}$ are added into the Fisher and then recalculated with the CRR influence. The '\planck only' parameters have been quoted by their marginalised values as well as their standard deviations.}
    \label{tab:alphaForecast}
\end{table}

\begin{table}
    \centering
    \begin{tabular}{c|c|c|c}
     \hline
     Parameters & \planck only & Voyage 2050  & \vtwentypp  \\
      & ($\mu_i \pm \sigma_i$) & \& \planck & \& \planck \\
     \hline\hline
       $\omega_b$ & $0.0199^{+0.0012}_{-0.0014}$ & $0.00044$ & $0.00001$ \\
       $\omega_c$ & $0.1058\pm0.0076$ & $0.0029$ & $0.0007$ \\
       $\me$ & $0.888\pm0.059$ & $0.020$ & $0.0026$ \\
       $100\, \theta_{\rm MC}$ & $0.958\pm0.045$ & $0.0155$ & $0.0022$ \\
       $\tau$ & $0.0512\pm0.0077$ & $0.0074$ & $0.0069$ \\
       $\ln(10^{10} A_s)$ & $3.029\pm0.017$ & $0.015$ & $0.014$ \\
       $n_s$ & $0.9640\pm0.0040$ & $0.0040$ & $0.0034$ \\
       \hline\hline
    \end{tabular}
    \caption{Similar error forecasts for $\me$ following the format of Table~\ref{tab:alphaForecast}. As before, the configurations are the same and the 'derived' parameters from \planck are not directly probed, rather they are modified by the CRR impact.}
    \label{tab:meForecast}
\end{table}

\changeJ{We also mention that in our forecasts we have not allowed for time-dependent variations of the fundamental constants. This had interesting effects on the CMB anisotropies and could be independently constrained \citep{Hart2017, Hart2021}. Here, we have three emission eras from each of the individual atomic species \citep{Sunyaev2009}. This means one could expect the interesting inter-atomic interplay to manifest in the CRR and even be constrainable with future CMB spectrometers. However, a more detailed analysis of this problem is beyond the scope of this paper.}

\section{Conclusions}\label{sec:conc}
In this work, we studied the effects of EDE and VFC on the CRR. We illustrated how the various model parameters affect the CRR and provided simple forecasts for the expected sensitivities of various CMB spectrometer concepts. 

The effect of EDE solely enters through changes of the expansion rate in the pre-recombination era. The associated effects on the three recombination phases depend on the details of the underlying model-parameters. For radiation-like dilution, all three recombination contributions can be modified, while for significantly faster dilution, the responses remain more localized in redshift. \changeJ{These effects in principle allow probing EDE models with spectrometers comparable to \vtwentypp. Since we do not have access to chains from \Planck for EDE model, we have not explored how a combination with a spectrometer could improve the constraints. However, we do anticipate significant gains but leave a more detailed analysis to future work.}

For VFC models, the recombination physics is directly affected, leaving distinct responses in the CRR that in principle again allow testing various phases in the pre-recombination era. The leading order effects for variations of $\aEM$ are a change in the amplitude of the CRR and smaller shift in the position of the recombination lines. The latter effect appears degenerate to changes in the value of the CMB monopole temperature; however, as we show here, the responses are distinct, in principle allowing to distinguish the two (see Fig.~\ref{fig:vfcCrrTotal}) directly with the CRR.
For variations of $\me$, we find the responses to be much smaller with significant cancellations between various effects (see lower panel of Fig.~\ref{fig:vfcCrrTotal}). 

\changeJ{Our simple forecasts show that the CRR provides the principle possibility to test EDE and VFC models. However, futuristic spectrometer sensitivities are required to derive independent but competitive constraints. By combining with CMB anisotropy measurements, significant improvements can be found. For example, \vtwenty with \Planck could improve the allowed error on variations of $\me$ by a factor of more than $\simeq 3$ and also remove a large part of the geometric degeneracy allowed by \Planck alone (see Fig.~\ref{fig:meFullContour}). This might shed new light on the origin of the Hubble tension, allowing us to rule out a VFC cause.}

\changeJ{A more comprehensive forecast, that combines the CRR responses with other cosmological probes could improve the forecasts, potentially allowing to distinguish between various scenarios. 
The addition of foregrounds will also be important. However, in contrast to $\mu$ and $y$ type distortions \citep{Abitbol2017}, the CRR does not suffer as strongly from foregrounds \citep{Hart2020c}, such that the main conclusions should not change as much.
In addition, EDE models can simultaneously create VFC effects, which then potentially enhance the sensitivity of the CRR to the underlying physics model.
An exploration of these possibilities is left to future work.}

\section{Data Availability}\label{sec:data}
The recombination lines were simulated and modelled using {\tt CosmoSpec} along with modifications for non-standard physics. The modelling of the Fisher matrix was created with the repository {\tt vfcFisher}, which will be made public upon the publication of this article\footnote{\url{http://www.github.com/cosmologyluke/vfcFisher}}.

\section*{Acknowledgements}
The authors would like to thank Alan Heavens for useful discussions surrounding the philosophy of Fisher matrices and likelihoods. All contour plots in this paper were made using the computing package {\tt GetDist}. This work was supported by the ERC Consolidator Grant CMBSPEC (No. 725456) as part of the European Union’s Horizon 2020 research and innovation program. LH would like to acknowledge the support of TNEI Services Ltd.

\bibliographystyle{mn2e}
\bibliography{Lit}
\appendix

\bsp	
\label{lastpage}
\end{document}